\documentclass[11pt]{article}

\usepackage{subcaption}

\usepackage{xcolor}
\usepackage{microtype}
\usepackage{url}            
\usepackage{amsfonts}       
\usepackage{nicefrac}       
\usepackage{mathtools, nccmath}
\usepackage{minitoc}
\usepackage{cancel}
\definecolor{darkgreen}{RGB}{0, 153, 0}

\usepackage{xcolor}

\newcommand{\yellowbox}[2]{
  \vspace{10pt} 
  \begingroup
  \setlength{\fboxsep}{10pt} 
  \setlength{\fboxrule}{2pt} 
  \noindent\fcolorbox{yellow!75!black}{yellow!5!white}{%
    \begin{minipage}{\dimexpr\linewidth-2\fboxsep-2\fboxrule\relax}
      \textbf{#1}\\[10pt] 
      #2 
    \end{minipage}%
  }
  \endgroup
  \vspace{10pt} 
}

\usepackage{microtype}
\usepackage{graphicx}
\usepackage{booktabs} 

\usepackage{comment}
\usepackage{url}
\usepackage{pythonhighlight}

\usepackage{amsfonts}       
\usepackage{amsmath}
\usepackage{amsfonts}
\usepackage{comment}

\usepackage{amsthm}
\usepackage{amscd}
\usepackage{t1enc}
\usepackage{enumerate}
\usepackage{indentfirst}
\usepackage{listings}
\usepackage{graphicx}
\usepackage{graphics}
\usepackage{pict2e}
\usepackage{epic}
\usepackage{epstopdf} 
\usepackage{listings}
\usepackage{minitoc}

\renewcommand{\nu}{\vartheta}

\usepackage{pgfplots} 

\usepackage{bm}
\usepackage{wrapfig}

\newtheorem*{theorem*}{Theorem}

\theoremstyle{definition}
\usepackage{algorithm}
\usepackage{algorithmic}

\usepackage{environ}
\usepackage{tikz}
\usepackage{float}

\newcounter{todos}

\usepackage{float}

\usepackage{xspace}
\newcommand{\coderec}{CodeRec\xspace}
\newcommand{\copilot}{Copilot\xspace}

\newcommand{\ghcp}{GiHub Copilot\xspace}

\newcommand{\cupsfull}{\coderec User Programming States\xspace}
\newcommand{\cups}{CUPS\xspace}
\newcommand{\sdevfirst}[1]{sample standard deviation, $s_N = #1$}
\newcommand{\sdev}[1]{($s_N = #1$)}  

\usepackage{fontawesome}

\usepackage{paralist}

\usepackage{blindtext}

\usepackage[left=1in, right =1in, top=1in, bottom=0.7in]{geometry}

\pdfoutput=1
\usepackage{booktabs}
\usepackage{pgfplots}
\usepackage{graphicx}
\usepackage{tikz}
\usepackage{amsmath, amsthm}
\usepackage{amssymb}
\usepackage{graphicx}
\usepackage{setspace}
\usepackage{xfrac}
\usepackage{xstring}
\usepackage{xspace}
\usepackage{bm}
\usepackage{microtype}
\usepackage{graphicx}

\usepackage{booktabs} 

\usepackage{comment}
\pgfplotsset{compat=1.14}

\usepackage{gensymb}
\usepackage{amsfonts}       
\usepackage{nicefrac}       
\usepackage{microtype}      
\usepackage{amsmath}
\usepackage{amsfonts}
\usepackage{amssymb}
\usepackage{comment}
\usepackage[hidelinks]{hyperref}
\usepackage{amsthm}
\usepackage{amscd}
\usepackage{t1enc}
\usepackage{enumerate}
\usepackage{enumitem}
\usepackage[mathscr]{eucal}
\usepackage{indentfirst}
\usepackage{listings}
\usepackage{graphicx}
\usepackage{graphics}
\usepackage{pict2e}
\usepackage{epic}
\usepackage{epstopdf} 
\usepackage{wrapfig}
\usepackage{algorithm}
\usepackage{algorithmic}
\usepackage{amsthm}
\usepackage{tabularx}
\usepackage{float}
\usepackage{setspace}
\usepackage{natbib}
\usepackage{fancyhdr}
\usepackage{paralist}
\usepackage{soul}
\usepackage{xcolor}

\definecolor{highlighter}{HTML}{fff100}
\sethlcolor{highlighter}
 \newcommand{\revision}[1]{#1}    
\usepackage[most]{tcolorbox}
\title{\textbf{Reading Between the Lines: Modeling User Behavior and Costs in AI-Assisted Programming}}
\allowdisplaybreaks
\usepackage{authblk}

\renewcommand{\headrulewidth}{0pt}

\author[1]{Hussein Mozannar}
\author[2]{Gagan Bansal}
\author[2]{Adam Fourney}
\author[2]{Eric Horvitz}
\affil[1]{Massachusetts Institute of Technology, Cambridge, USA}
\affil[2]{Microsoft Research, Redmond, USA}
\date{}
\begin{document}
\newgeometry{left=1in, right =1in, top=0.6in, bottom=0.7in}

\begin{titlepage}
\maketitle
\vspace{-1cm}
\begin{abstract}
Code-recommendation systems, such as Copilot and CodeWhisperer, have the potential to improve programmer productivity by suggesting and auto-completing code. However, to fully realize their potential, we must understand how programmers interact with these systems and identify ways to improve that interaction. To seek insights about human-AI collaboration with code recommendations systems, we studied GitHub Copilot, a code-recommendation system used by millions of programmers daily. 
We developed CUPS, a taxonomy of common programmer activities when interacting with Copilot. Our study of 21 programmers, who completed coding tasks and retrospectively labeled their sessions with CUPS, showed that CUPS can help us understand how programmers interact with code-recommendation systems, revealing inefficiencies and time costs. Our insights reveal how programmers interact with Copilot and motivate new interface designs and metrics. 

\end{abstract}

\begin{figure}[H]
\centering
  \includegraphics[scale=0.35,trim={0cm 0cm 0 0}]{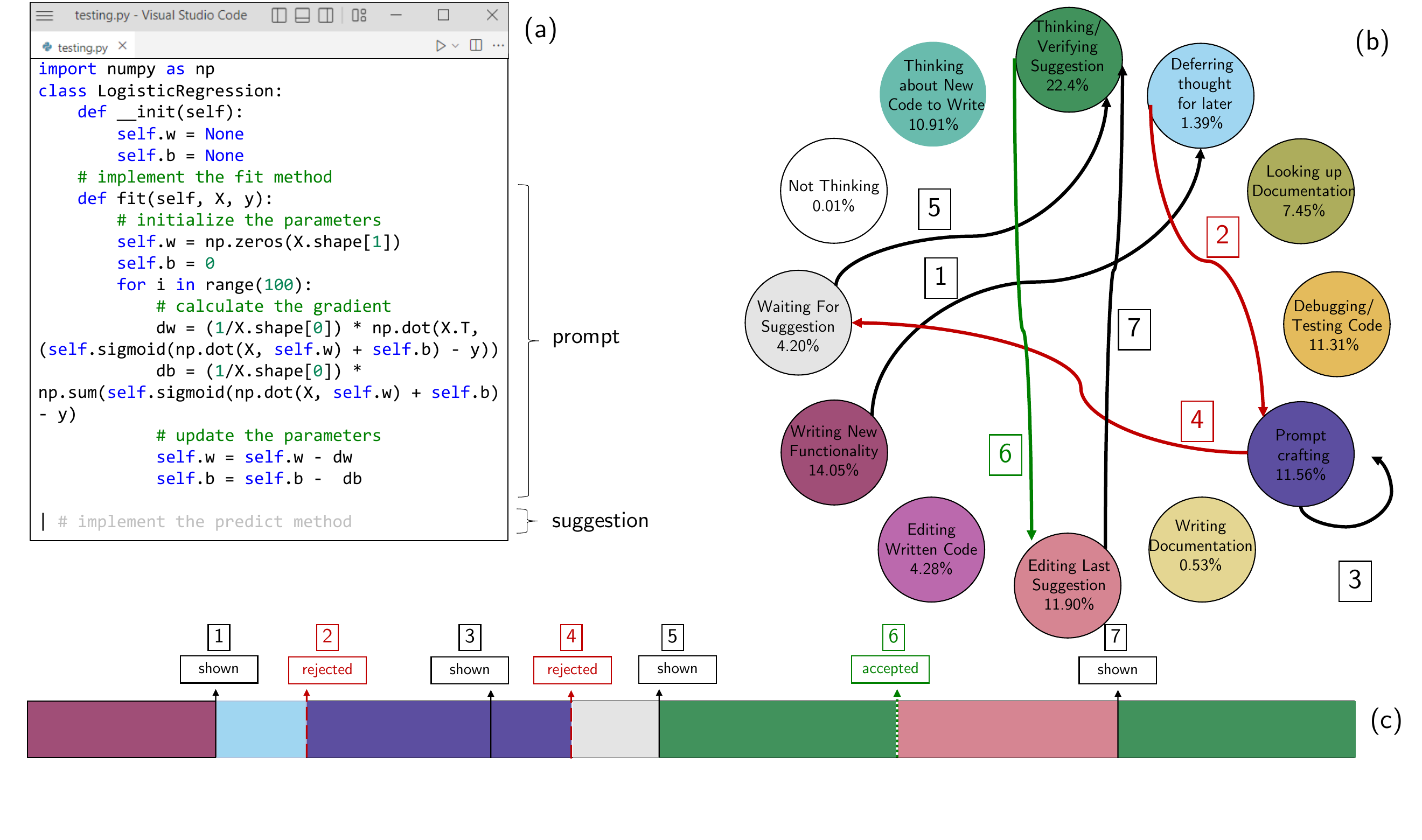}
  \caption{Profiling a coding session with the \cupsfull (\cups). In (a) we show the operating mode of \coderec inside Visual Studio Code. In (b) we show the  \cups taxonomy used to describe \coderec related programmer activities. A coding session  can  be summarized as a timeline in (c) where the programmer transitions between states.}
  \label{fig:figure1}
\end{figure}

\end{titlepage}
\restoregeometry

\section{Introduction}

Programming-assistance systems based on the adaptation of large language models (LLMs) to code recommendations have been recently introduced to the public.
Popular systems, including Copilot \cite{copilot}, CodeWhisperer \cite{codwhisperer}, and AlphaCode\cite{li2022competition}, signal a potential shift in how software is developed.
Though there are differences in specific interaction mechanisms, the programming-assistance systems generally extend existing IDE code completion mechanisms (e.g., IntelliSense~\footnote{\url{https://code.visualstudio.com/docs/editor/intellisense}}) by producing suggestions using neural models trained on billions of lines of code \cite{chen2021evaluating}.
The LLM-based completion models can suggest sentence-level completions to entire functions and classes in a wide array of programming languages.
These large neural models are deployed with the goal of accelerating the efforts of software engineers, reducing their workloads, and improving their productivity.

Early assessments suggest that programmers do feel more productive when assisted by the code recommendation models~\cite{ziegler2022productivity} and that they prefer these systems to earlier code completion engines~\cite{vaithilingam2022expectation}.
In fact, a recent study from GitHub, 
found that Copilot could potentially reduce task completion time by a factor of two {\cite{peng2023impact}}.
While these studies help us understand the benefits of code-recommendation systems, they do not allow us to identify avenues to improve and understand the nature of interaction with these systems.

In particular, the neural models introduce new tasks into a developer's workflow, such as writing AI prompts~\cite{jiang2022discovering} and verifying AI suggestions~\cite{vaithilingam2022expectation}, which can be lengthy.
Existing interaction metrics, such as suggestion acceptance rates, time to accept (i.e., the time a suggestion remains onscreen), and reduction of tokens typed, tell only part of this interaction story.
For example, when suggestions are presented in monochrome popups (Figure \ref{fig:figure1}), programmers may choose to accept them into their codebases so that they can be read with code highlighting enabled.
Likewise, when models suggest only one line of code at a time, programmers may accept sequences before evaluating them together as a unit.
In both scenarios, considerable work verifying and editing suggestions occurs \emph{after} the programmer has accepted the recommended code. Prior interaction metrics also largely miss user effort invested in devising and refining prompts used to query the models.
When code completion tools are evaluated using coarser task-level metrics such as task completion time~\cite{kalliamvakou_2022}, we begin to see signals of the benefits of AI-driven code completion but lack sufficient detail to understand the nature of these gains, as well as possible remaining inefficiencies.
We argue that an ideal approach would be sufficiently low level to support interaction profiling while sufficiently high level to capture meaningful programmer activities.

Given the nascent nature of these systems, numerous questions exist regarding the behavior of their users:
\begin{itemize}
\item What activities do users undertake in anticipation for, or to trigger a suggestion?
\item  What mental processes occur while the suggestions are onscreen, and, do people double-check suggestions before or after acceptance?
\item 
How {\em costly} for users are these various new tasks, and which take the most time?
\end{itemize}

To answer these and related questions in a systematic manner, we apply a mixed-methods approach to analyze interactions with a popular code suggestion model, \ghcp \footnote{\url{https://github.com/features/copilot}} which has more than a million users. To emphasize that our analysis is not restricted to the specifics of \copilot, we use the term \coderec to refer to any instance of code suggestion models, including \copilot.
 Through small-scale pilot studies and our first-hand experience using \copilot for development, we develop a novel taxonomy of common states of a programmer when interacting with \coderec models (such as \copilot), which we refer to as \cupsfull (\cups). The \cups taxonomy serves as the main tool to answer our research questions.

Given the initial taxonomy, we conducted a user study with 21 developers who were asked to retrospectively review videos of their coding sessions and explicitly label their intents and actions using this model, with an option to add new states if necessary.
The study participants labeled a total of 3137 coding segments and interacted with 1096 suggestions.
The study confirmed that the taxonomy was sufficiently expressive, and we further learned transition weights and state dwell times ---something we could not do without this experimental setting.
Together, these data can be assembled into various instruments, such as the \cups diagram (Figure \ref{fig:figure1}), to facilitate profiling interactions and identify inefficiencies.
Moreover, we show that such analysis nearly doubles our estimates for how much developer time can be attributed to interactions with code suggestion systems, as compared with existing metrics.
We believe that identifying the current CUPS state during a programming session can help serve programmer needs.
This can be accomplished using custom keyboard macros or automated prediction of CUPS states, as discussed in our future work section and the Appendix.
Overall, we leverage the \cups diagram to identify some opportunities to address inefficiencies in the current version of \copilot.

In sum, our main contributions are the following: 
\begin{itemize}
    \item A novel taxonomy of common activities of programmers (called \cups) when interacting with code recommendation systems (Section \ref{sec:cups})
    \item A dataset of coding sessions annotated with user actions, \cups, and video recordings of programmers coding with \copilot (Section \ref{sec:user_study}).
    \item Analysis of which \cups states programmers spend their time in when completing coding tasks (Subsection \ref{subsec:time_in_cups}).
    \item An instrument to analyze programmer behavior (and patterns in behavior) based on a finite-state machine on \cups states (Subsection \ref{subsec:patterns}).
    \item An adjustment formula to properly account for how much time do programmers spend verifying \coderec suggestions (Subsection \ref{subsec:time_adjustment}) inspired by the \cups state of deferring thought (Subsection \ref{subsec:defer_thought}).
\end{itemize}

The remainder of this paper is structured as follows: We first review related work on AI-assisted programming (Section \ref{sec:related_work}) and formally describe \copilot, along with a high-level overview of programmer-\coderec interaction (Section \ref{sec:setup}). To further understand this interaction, we define our model of \cupsfull (\cups) (Section \ref{sec:setup}) and then describe a user study designed to collect programmer annotations of their states (Section \ref{sec:user_study}). We use the collected data to analyze the interactions using \cups diagram revealing new insights into programmer behavior (Section {\ref{sec:study_results})}. We then discuss limitations and future work and conclude in  (Section \ref{sec:discussion}).

\section{Background and Related Work}\label{sec:related_work}

Large language models based on the Transformer network \cite{vaswani2017attention}, such as GPT-3 \cite{brown2020language}, have found numerous applications in natural language processing. Codex \cite{chen2021evaluating}, a GPT model trained on 54 million GitHub repositories, demonstrates that LLMs can very effectively solve various programming tasks. Specifically, Codex was initially tested on the HumanEval dataset containing 164 programming problems, where it is asked to write the function body from a docstring \cite{chen2021evaluating} and achieves 37.7\% accuracy with a single generation. Various metrics and datasets have been proposed to measure the performance of code generation models \cite{hendrycks2021measuring,li2022competition,evtikhiev2022out, dakhel2022github}. However, in each case, these metrics test how well the model can complete code in an offline setting without developer input rather than evaluating how well such recommendations assist programmers in situ. 
This issue has also been noted in earlier work on non-LLM based code completion models where performance on completion benchmarks overestimates the model's utility to developers \cite{hellendoorn2019code}. Importantly, however, these results may not hold to LLM-based approaches, which are radically different \cite{sarkar2022like}.

One straightforward approach to understanding the utility of neural code completion services, including their propensity to deliver incomplete or imperfect suggestions, is to simply ask developers. To this end, Weisz et al. interviewed developers and found that they did not require a perfect recommendation model for the model to be useful \cite{weisz2021perfection}. Likewise, Ziegler et al. surveyed over 2,000  Copilot users \cite{ziegler2022productivity} and asked about perceived productivity gains using a survey instrument based on the SPACE framework \cite{forsgren2021space} -- we incorporate the same survey design for our own study. 
They found both that developers felt more productive using Copilot and that these self-reported perceptions were reasonably correlated with suggestion acceptance rates. \citet{liang2023understanding} administered a survey to 410 programmers who use various AI programming assistants, including Copilot, and highlighted why the programmers use the AI assistants and numerous usability issues. Similarly, \citet{prather2023s} surveyed how introductory programming students utilize Copilot. 

While these self-reported measures of utility and preference are promising, we would expect gains to be reflected in objective metrics of productivity. Indeed, one ideal method would be to conduct randomized control trials where one set of participants writes code with a recommendation engine while another set codes without it. GitHub performed such an experiment where 95 participants were split into two groups and asked to write a web server. The study concluded by finding that task completion was reduced by 55.8\% in the Copilot condition \cite{peng2023impact}. Likewise, a study by Google showed that an internal \coderec model had a 6\% reduction in 'coding iteration time' \cite{google_ai_blog_2022}. On the other hand, \citet{vaithilingam2022expectation} showed in a study of 24 participants showed no significant improvement in task completion time -- yet participants stated a clear preference for Copilot. An interesting comparison to Copilot is Human-Human pair programming, which \citet{wu2023ai} details.

\revision{
A significant amount of work has tried to understand the behavior of programmers}{\cite{brooks1980studying,brooks1977towards,sheil1981psychological,lieberman1995bridging}}\revision{ using structured user studies under the name of "psychology of programming." This line of work tries to understand the effect of programming tools on the time to solve a task or ease of writing code and how programmers read and write code. Researchers often use telemetry with detailed logging on keystrokes} {\cite{velart2006user,ju2018teamscope}} \revision{to understand behavior. Moreover, eye-tracking is also used to understand how programmers read code}{\cite{peitek2020drives,obaidellah2018survey}}.\revision{ Our research uses raw telemetry alongside user-labeled states to understand behavior; future research could also utilize eye-tracking and raw video to get deeper insights into behavior.
}

This wide dispersion of results raises interesting questions about the nature of the utility afforded by neural code completion engines: how, and when, are such systems most helpful; and conversely, when do they add additional overhead? \textbf{This is the central question to our work.} The related work closest to answering this question is that of Barke et al. \cite{barke2022grounded}, who showed that interaction with Copilot falls into two broad categories: the programmer is either in ``acceleration mode'' where they know what they want to do, and Copilot serves to make them faster; or they are in ``exploration mode'', where they are unsure what code to write and Copilot helps them explore. The taxonomy we present in this paper, CUPS, enriches this further with granular labels for programmers' intents. Moreover, the data collected in this work was labeled by the participants themselves rather than by the researchers interpreting their actions, allowing for more faithful intent and activity labeling and the data collected in our study can also be used to build predictive models as in \cite{sun2023don}. 
The next section describes the \copilot system formally and describes the data collected when interacting with \copilot.

\section{\copilot System Description}\label{sec:setup}

  \begin{figure}[ht]
     \centering
     \includegraphics[width=0.49\textwidth]{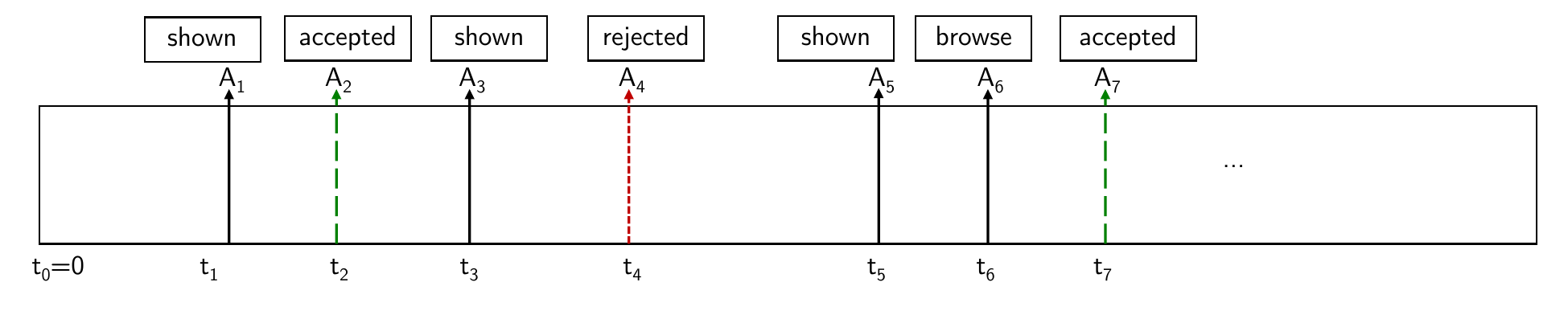}
     \caption{Schematic of interaction telemetry with \copilot as a timeline. For a given coding session, the telemetry contains a sequence of timestamps and actions with associated prompt and suggestion features (not shown).}
     \label{fig:telemetry}

 \end{figure}

To better understand how code recommendation systems influence the effort of programming, we focus on \ghcp, a popular and representative example of this class of tools. \copilot \footnote{The version of Copilot that this manuscript refers to is Copilot as of August 2022.} is based on a Large Language Model (LLM) and assists programmers inside an IDE by recommending code suggestions any time the programmer pauses their typing.  Figure~\ref{fig:figure1} shows an example of \copilot recommending a code snippet as an inline, monochrome popup, which the programmer can accept using a keyboard shortcut (e.g., <tab>).

To serve suggestions, \copilot uses a portion of the code written so far as a {\em prompt}, $P$, which it passes to the underlying LLM. The model then generates a suggestion, $S$, which it deems to be a likely completion. In this regime, programmers can \emph{engineer} the prompt to generate better suggestions by carefully authoring natural language comments in the code such as ``\texttt{\# split the data into train and test sets}.'' 
In response to a \copilot suggestion, the programmer can then take one of several actions $A$, where $A \in \{\textrm{browse, accept, reject}\}$. The latter of these actions, \emph{reject}, is triggered implicitly by continuing to type \revision{something that differs from the suggestion or by pressing the escape key. The browse action enables the programmer to change the suggestion shown with a keyboard shortcut from a set of at most three suggestions}.  
\copilot logs aspects of the interactions via {\em telemetry}. We leverage this telemetry in the studies described in this paper.
Specifically, whenever a suggestion is shown, accepted, rejected, or browsed, we record a tuple to the telemetry database, $(t_i, A_i, P_i, S_i)$, where $t_i$ represents the within-session timestamp of the $i^\text{th}$ event ($t_0 = 0$), $A_i$ details the action taken (augmented to include `shown'), and $P_i$ and $S_i$ capture features of the prompt and suggestion, respectively. Figure~\ref{fig:telemetry} displays telemetry of a coding session, and Figure~\ref{fig:figure1}a shows \copilot implemented as a VSCode plugin. We have the ability to capture telemetry for any programmer interacting with {\copilot}; this is used to collect data for a user study in section {\ref{sec:user_study}}.

\subsection{Influences of \coderec on Programmer's Activities} 
Despite the limited changes that \copilot introduces to an IDE's repertoire of actions, LLM-based code suggestions can significantly influence how programmers author code.
Specifically, \copilot leverages LLMs to stochastically {\em generate} novel code to fit the arbitrary current context. As such, the suggestions may contain errors (and can appear to be unpredictable)
and require that programmers double-check and edit them for correctness.
Furthermore, programmers may have to refine the prompts to get the best suggestions.
These novel activities associated with the AI system introduce new efforts and potential disruptions to the flow of programming. 
{\bf We use time as a proxy to study the new costs of interaction introduced by the AI system.}
We recognize that this approach is incomplete: the costs associated with solving programming tasks are multi-dimensional, and it can be challenging to assign a single real-valued number to cover all facets of the task~\cite{forsgren-cacm-2021}. 
Nevertheless, we argue that, like accuracy, efficiency-capturing measures of time are an important dimension of the cost that is relevant to most programmers.

\subsection{Programmer Activities in Telemetry Segments}

\copilot's telemetry captures only instantaneous user actions (e.g., accept, reject, browser), as well as the suggestion display event. By themselves, these entries do not reveal such programmer's activities as double-checking and prompt engineering, as such activities happen \emph{between} two consecutive instantaneous events. 
{\bf We argue that the regions between events, which we refer to as {\em telemetry segments}, contain important user intentions and activities unique to programmer-\coderec interaction},
which we need to understand in order to answer how \copilot affects programmers---and where and when \copilot suggestions are useful to programmers.

Building on this idea, telemetry segments can be split into two groups (Figure \ref{fig:telemetry}). The first group includes segments that start with a suggestion shown event and end with an action (accept, reject, or browse). Here, the programmer is paused and has yet to take action. We refer to this as `User Before Action'. The second group includes segments that start with an action event and end with a display event. During this period, the programmer can be either typing or paused; hence we denote it as `User Typing or Paused'. 
These two groups form the foundation of a deeper taxonomy of programmers' activities, which we will further develop in the next section.

\section{A Taxonomy for Understanding Programmer-\coderec Interaction: \cups}\label{sec:cups}

 \begin{figure*}[t]
     \centering
     \includegraphics[width=\textwidth, scale=0.6]{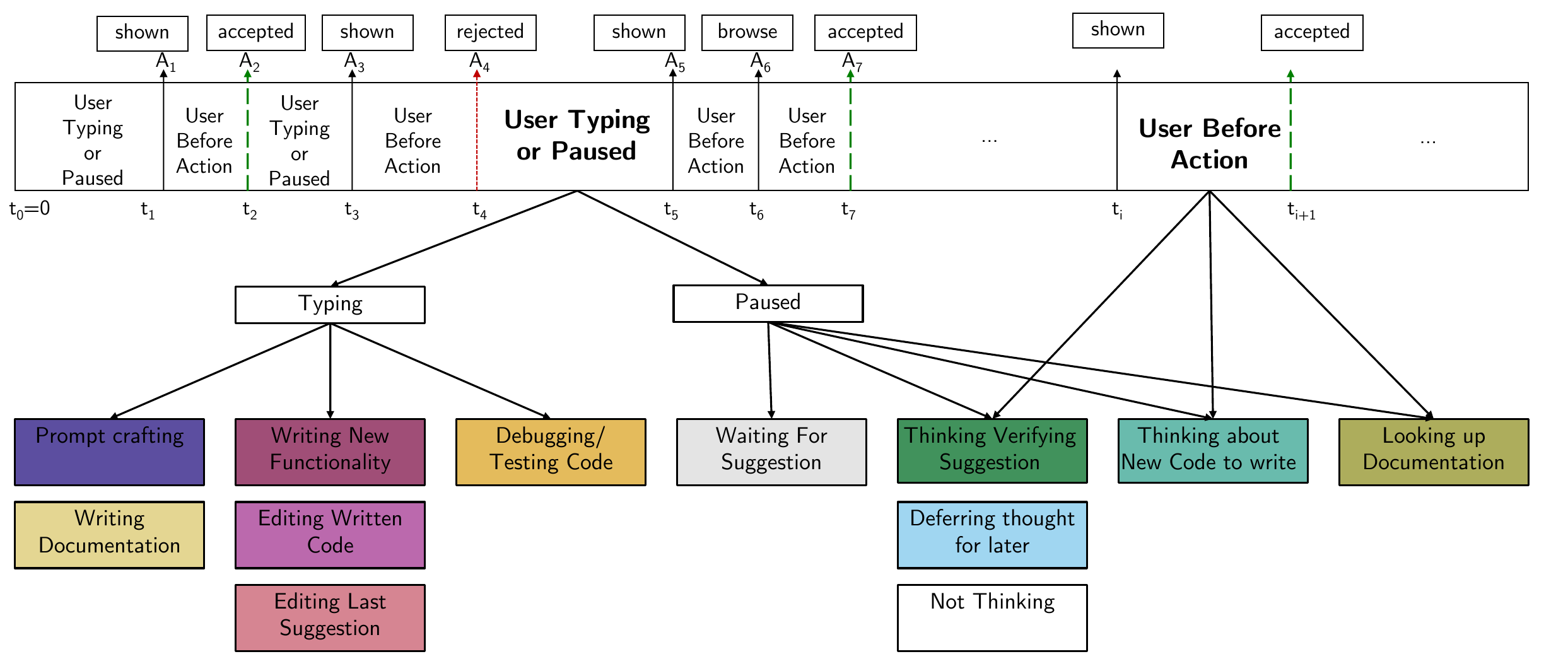}
     \caption{Taxonomy of programmer's activities when interacting with \coderec -- \cups. }
     \label{fig:cups}   
 \end{figure*}

\begin{table}[h]
  \caption{Description of each state in \cupsfull (\cups). }
  \centering
 \begin{tabular}{p{0.35\textwidth}p{0.6\textwidth}}
    \toprule
    \textbf{State} & \textbf{Description} \\
    \midrule
Thinking/Verifying Suggestion & Actively thinking about and verifying a shown or accepted suggestion \\ 
Not Thinking  & Not thinking about suggestion or code, programmer away from keyboard \\
Deferring Thought For Later & Programmer accepts suggestion without completely verifying it, but plans to verify it after  \\
Thinking About New Code To Write & Thinking about what code or functionality to implement and write \\
Waiting For Suggestion & Waiting for \coderec suggestion to be shown   \\
Writing New Code & Writing code that implements new functionality \\
Editing Last Suggestion & Editing the last accepted  suggestion \\
Editing (Personally) Written Code& Editing code written by a programmer that is not a \coderec suggestion for the purpose of fixing existing functionality \\
Prompt Crafting & Writing prompt in the form of comment or code to obtain desired \coderec suggestion\\
Writing Documentation & Writing comments or docstring for purpose of documentation\\
Debugging/Testing Code & Running or debugging code to check functionality may include writing tests or debugging statements\\
Looking up Documentation & Checking an external source for the purpose of understanding code functionality (e.g. Stack Overflow)  \\
Accepted & Accepted a \coderec suggestion \\
Rejected & Rejected a \coderec suggestion \\
  \bottomrule
\end{tabular}%

  \label{tab:cups_description}

\end{table} 

\subsection{Creating the Taxonomy}
\label{sec:labeltool}
\begin{figure*}[t]
    \centering
    \includegraphics[width=\textwidth, scale=0.35]{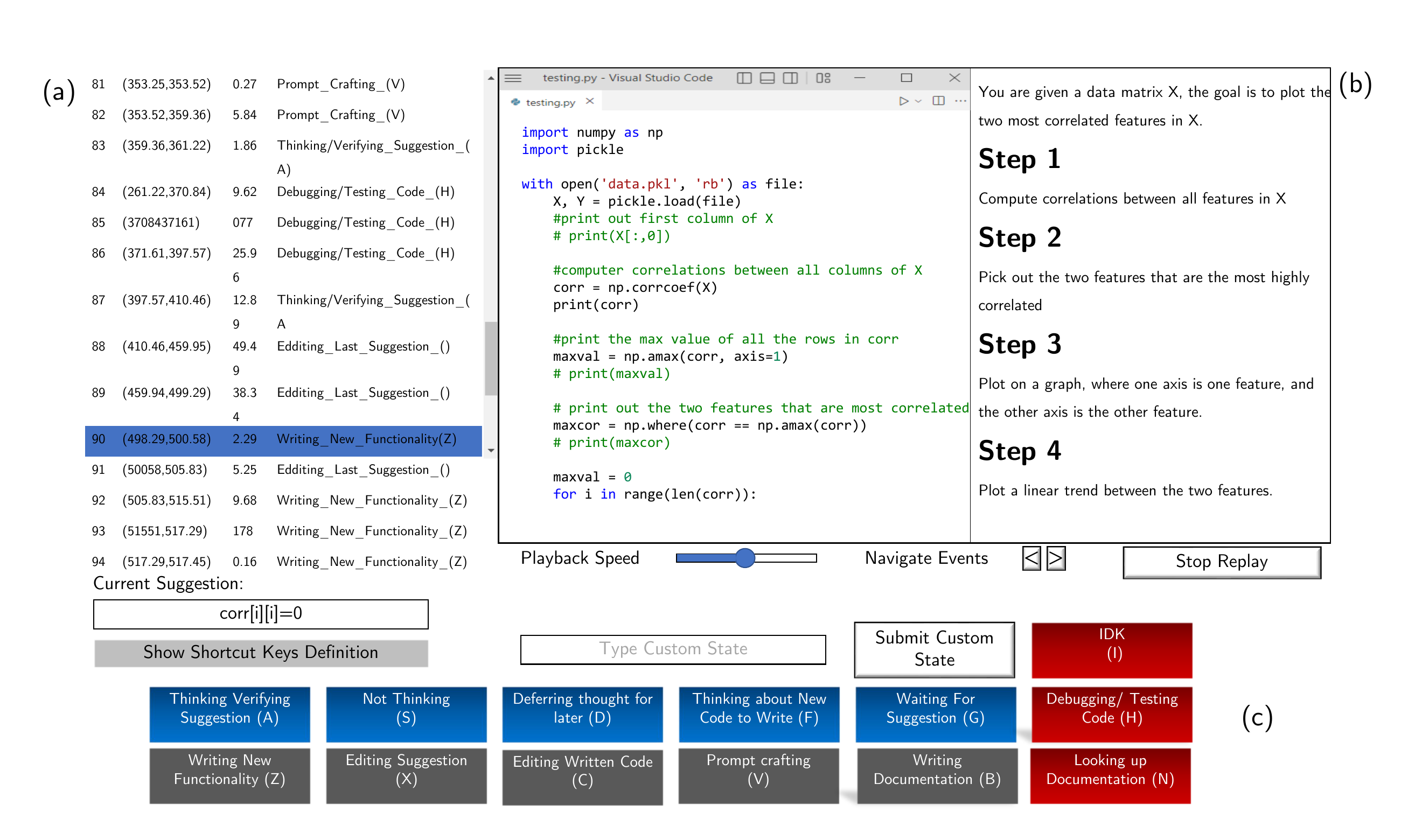}
    \caption{Screenshot of retrospective labeling tool for coding sessions. Left: Navigation panel for telemetry segments. Right: Video player for reviewing video of a coding session. Bottom: Buttons and text box for labeling states.}
    \label{fig:interface}
\end{figure*}

Our objective is to create an extensive, but not complete, taxonomy of programmer activities when interacting with {\coderec} that enables a useful study of the interaction.
To refine the taxonomy of programmers' activities, we developed a labeling tool and populated it with an initial set of activities based on our own experiences from extensive interactions with \copilot (Figure \ref{fig:interface}). The tool enables users to watch a recently captured screen recording of them solving a programming task with \copilot's assistance and to {\em retrospectively} annotate each telemetry segment with an activity label. We use this tool to first refine our taxonomy with a small pilot study (described below) and then to collect data in Section {\ref{sec:user_study}}.

The labeling tool (Figure~\ref{fig:interface}) contains three main sections: 
a) A navigation panel on the left, which displays and allows navigating between telemetry segments and highlights the current segment being labeled in blue. The mouse or arrow keys are used to navigate between segments.
b) A video player on the right, which plays the corresponding video segments in a loop. The participant can watch the video segments any number of times.
c) Buttons on the bottom corresponding to the CUPS taxonomy, along with an ``IDK'' button and a free-form text box to write custom state labels. Buttons also have associated keyboard bindings for easy annotation.

To label a particular video segment, we asked participants to consider the hierarchical structure of \cups in Figure \ref{fig:cups}.\revision{ The hierarchical structure first distinguishes segments by whether a typing segment occurred in that segment and then decides based on the typing or non-typing states.} For example, in a segment where a participant was initially double-checking a suggestion and then wrote new code to accomplish a task, the appropriate label would be "Writing New Functionality" as the user eventually typed in the segment.  In cases where there are two states that are appropriate and fall under the same hierarchy, e.g., if the participant double-checked a suggestion and then looked up documentation, they were asked to pick the state in which they spent the majority of the time. These issues arise because we collect a single state for each telemetry segment.

\textbf{Pilot.}
Through a series of pilots involving the authors of the paper, as well as three other participants drawn from our organization, we iteratively applied the tool to our own coding sessions and to the user study tasks described in section {\ref{sec:user_study}}. We then expanded and refined the taxonomy by incorporating any ``custom state'' (using the text field) written by the pilot participants. The states 'Debugging/Testing Code', 'Looking up Documentation', and 'Writing Documentation' were added through the pilots. By the last pilot participant, the code book was stable and saturated as they did not write a state that was not yet covered. We observed in our study that the custom text field was rarely used. We describe the resultant taxonomy in the sections below.

\subsection{Taxonomy of Telemetry Segments}

Figure~\ref{fig:cups} shows the finalized taxonomy of programmer activities for individual telemetry segments with \copilot. As noted earlier, the taxonomy is rooted in two segment types: `User Typing or Paused', and `User Before Action'. We first detail the `User Typing or Paused' segments, which precede shown events (Figure \ref{fig:telemetry}) and are distinguished by the fact that no suggestions are displayed during this time. As the name implies, users can find themselves in this state if they are either actively 'Typing'\footnote{Active typing allows for brief pauses between keystrokes.}, or have 'paused' but have not yet been presented with a suggestion. 
In cases where the programmer is actively typing, they could be completing any of a number of tasks such as: `writing new functionality,' 'editing existing code,' 'editing prior (\coderec) suggestions,' `debugging code,' or authoring natural language comments, including both documentation and prompts directed at \coderec (i.e., `prompt crafting'). 
When the user pauses, they may simply be ``waiting for a suggestion'' or can be in any number of states common to `User Before Action' segments.

In every `User Before Action' segment, \coderec is displaying a suggestion, and the programmer is paused and not typing. They could be reflecting and verifying that suggestion, or they may not be paying attention to the suggestion and thinking about other code to write instead. The programmer can also {\em defer} their efforts on the suggestion for a later time period by accepting it immediately, then pausing to review the code at a later time. This can occur, for example, because the programmer desires syntax highlighting rather than grey text or because the suggestion is incomplete, and the programmer wants to allow \copilot to complete its implementation before evaluating the code as a cohesive unit. The latter situation tends to arise when \copilot displays code suggestions line by line (e.g., Figure \ref{fig:defer_thought}). 

The leaf nodes of the finalized taxonomy represent 12 distinct states that programmers can find themselves in. These states are illustrated in Figure \ref{fig:cups} and are further described in Table \ref{tab:cups_description}. While the states are meant to be distinct, siblings may share many traits. For example, "Writing New Functionality" and "Editing Written Code" are conceptually very similar. This taxonomy also bears resemblance to the keystroke level model in that it assigns a time cost to mental processes as well as typing \cite{card1980keystroke,john1996goms}. As evidenced by the user study---which we describe in the next section---these 12 states provide a language that is both \emph{general} enough to capture most activities (at this level of abstraction), and \emph{specific} enough to meaningfully capture activities unique to LLM-based code suggestion systems.

\section{\cups Data Collection Study}
\label{sec:user_study}

To study \coderec-programmer interaction in terms of \cupsfull, we designed a user study where programmers perform a coding task, then review and label videos of their coding session using the telemetry segment-labeling tool described earlier. We describe the procedure, the participants, and the results in the sections that follow.

\subsection{Procedure}
\label{subsec:user_study_setup}

We conducted the study over a video call and asked participants to use a remote desktop application to access a virtual machine (VM). Upon connecting, participants were greeted with the study environment consisting of Windows 10, together with Visual Studio Code (VS Code) augmented with the \copilot plugin. 

Participants were then presented with a programming task drawn randomly from a set of eight pre-selected tasks (Table \ref{tab:coding_tasks}). \revision{If the participant was unfamiliar with the task content, we offered them a different random task.} The task set was designed during the pilot phase so that individual tasks fit within a 20-minute block and so that, together, the collection of tasks surfaces a sufficient diversity of programmer activities. It is crucial that the task is of reasonable duration so that participants are able to remember all their activities since they will be required to label their session immediately afterward. Since the {\cups} taxonomy includes states of thought, participants must label their session immediately after coding, and each study took approximately 60 minutes in total.
To further improve diversity, task instructions were presented to participants as images to encourage participants to author their own \copilot prompts rather than copying and pasting from the problem description. The full set of tasks and instructions is provided as an Appendix. 

Upon completing the task (or reaching the 20-minute mark), we loaded the participant's screen recording and telemetry into the labeling tool (previously detailed in Section~\ref{sec:labeltool}). The researcher then briefly demonstrated the correct operation of the tool and explained the CUPS taxonomy. Participants were then asked to annotate their coding session with CUPS labels. \revision{Self-labeling allows us to easily scale such a study and enables more accurate labels for each participant, but may cause inconsistent labeling across participants.} Critically, this labeling occurred within minutes of completing the programming task so as to ensure accurate recall.
We do not include a baseline condition where participants perform the coding task without {\copilot}, as this work focuses on understanding and modeling the interaction with the current version of {\copilot}.

Finally, participants completed a post-study questionnaire about their experience mimicking the one in \cite{ziegler2022productivity}. The entire experiment was designed to last 60 minutes. The study was approved by our institutional review board (IRB), and participants received a \$50.00 gift card as remuneration for their participation.

\begin{table*}[ht]
  \caption{Description of the coding tasks given to user study participants and task assignment. \revision{Participants were randomly allocated to tasks for tasks which they had familiarity with.}}
  \label{tab:coding_tasks}
 \resizebox{\textwidth}{!}{\begin{tabular}{lll}
    \toprule
    \textbf{Task Name} & \textbf{Participants} & \textbf{Description} \\
    \midrule
    Algorithmic Problem & P4,P17,P18 & Implementation of TwoSum, ThreeSum and FourSum  \\
    Data Manipulation & P1,P2,P11,P20 & Imputing data with average feature value and feature engineering for quadratic terms \\
    Data Analysis &P5,P8 & Computing data correlations in a matrix and plotting of most highly correlated features \\
    Machine Learning &P3,P7,P12,P15 & Training and Evaluation of models using sklearn on given dataset \\
    Classes and Boilerplate Code &P6,P9 & Creating different classes that build on each other \\
    Writing Tests &P16 & Writing tests for a black box function that checks if a string has valid formatting \\
    Editing Code &P10,P14,P21 & Adding functionality to an existing class that implements a nearest neighbor retriever \\
    Logistic Regression &P13,P19 & Implementing a custom Logistic Regression from scratch with weight regularization \\
  \bottomrule
\end{tabular}
}

\end{table*} 

\subsection{Participants} To recruit participants, we posted invitations to developer-focused email distribution lists within our large organization. We recruited 21 participants with varying degrees of experience using \copilot:  7 used \copilot more than a few times a week, 3 used it once a month or less, and 11 had never used it before. For participants who had never used it before, the experimenter gave a short oral tutorial on \copilot explaining how it can be invoked and how to accept suggestions. 
Participants' roles in the organization ranged from software engineers (with different levels of seniority) to researchers and graduate student interns.
In terms of programming expertise, only  6 participants had less than 2 years of professional programming experience (i.e., excluding years spent learning to program), 5 had between 3 to 5 years, 7 had between 6 to 10 years, and 3 had more than 11 years of experience. 
Participants used a language in which they stated proficiency (defined as language in which they were comfortable designing and implementing whole programs). Here, 19 of the 21 participants used Python, one used C++, and the final participant used JavaScript.

On average, participants took $12.23$ minutes (\sdevfirst{3.98} minutes) to complete the coding task, with a maximum session length of 20.80 minutes.
This task completion time is measured from the first line of code written for the task until the end of the allocated time.
During the coding tasks, \copilot showed participants a total of 1024 suggestions, out of which they accepted 34.0\%. The average acceptance rate for participants was 36.5\% (averaging over the acceptance rate of each participant), and the median was 33.8\% with a standard error of 11.9\%; the minimum acceptance rate was 14.3\%, and the maximum was 60.7\%.  In the labeling phase, each participant labeled an average of $149.38$ \sdev{57.43} segments with \cups, resulting in a total of 3137 \cups labels. 
The participants used the `custom state' text field only three times total, twice a participant wrote `write a few letters and expect suggestion' which can be considered as `prompt crafting' and once a participant wrote `I was expecting the function skeleton to show up[..]' which was mapped to 'waiting for suggestion'. The IDK button was used a total of 353 times, this sums to 3137 {\cups} + 353 IDKs = 3490 labels, the majority of its use was from two participants (244 times) where the video recording was not clear enough during consecutive spans, and was used by only five other participants more than once with the majority of the use also being due to the video not being clear or the segment being too short. \revision{The IDK segments represent 6.5\%  of total session time across all participants, mostly contributed by five participants. Therefore,} we remove the IDK segments from the analysis and do not attempt to re-label them.

Together, these \cups labels enable us to investigate various questions about programmer-\coderec interaction systematically, such as exploring
which activities programmers perform most frequently and how they spend most of their time. We study programmer-\coderec interaction using the data derived from this study in the following Section~\ref{sec:study_results} and derive various insights and interventions. 

 \begin{figure*}
   \begin{subfigure}[b]{1.0\textwidth}
    \includegraphics[width=\textwidth, height = 7cm]{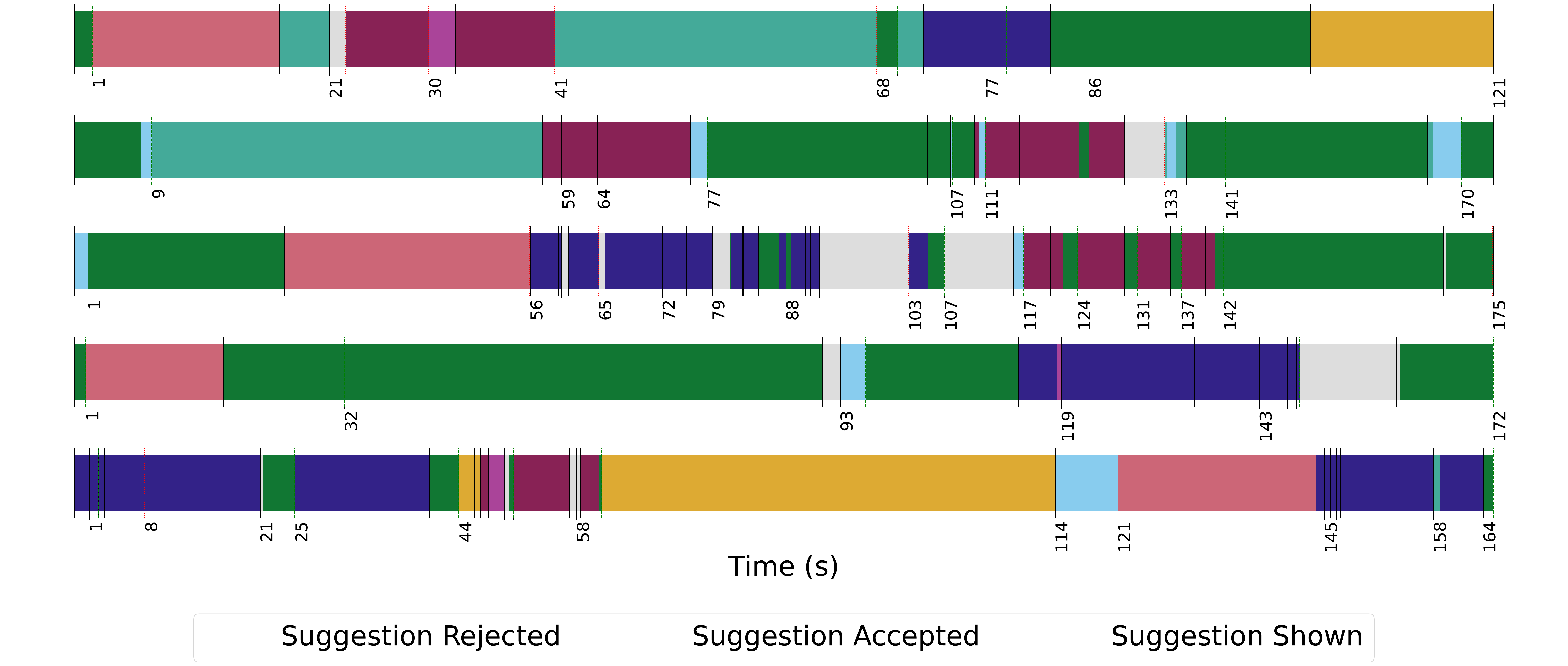}
    \caption{Individual \cups timelines for 5/21 study participants for the first 180 secs show the richness of and variance in programmer-\coderec interaction.}
    \label{fig:timelines}
  \end{subfigure}
  \hfill
  \begin{subfigure}[b]{0.40\textwidth}
    \includegraphics[angle=90,width=\textwidth]{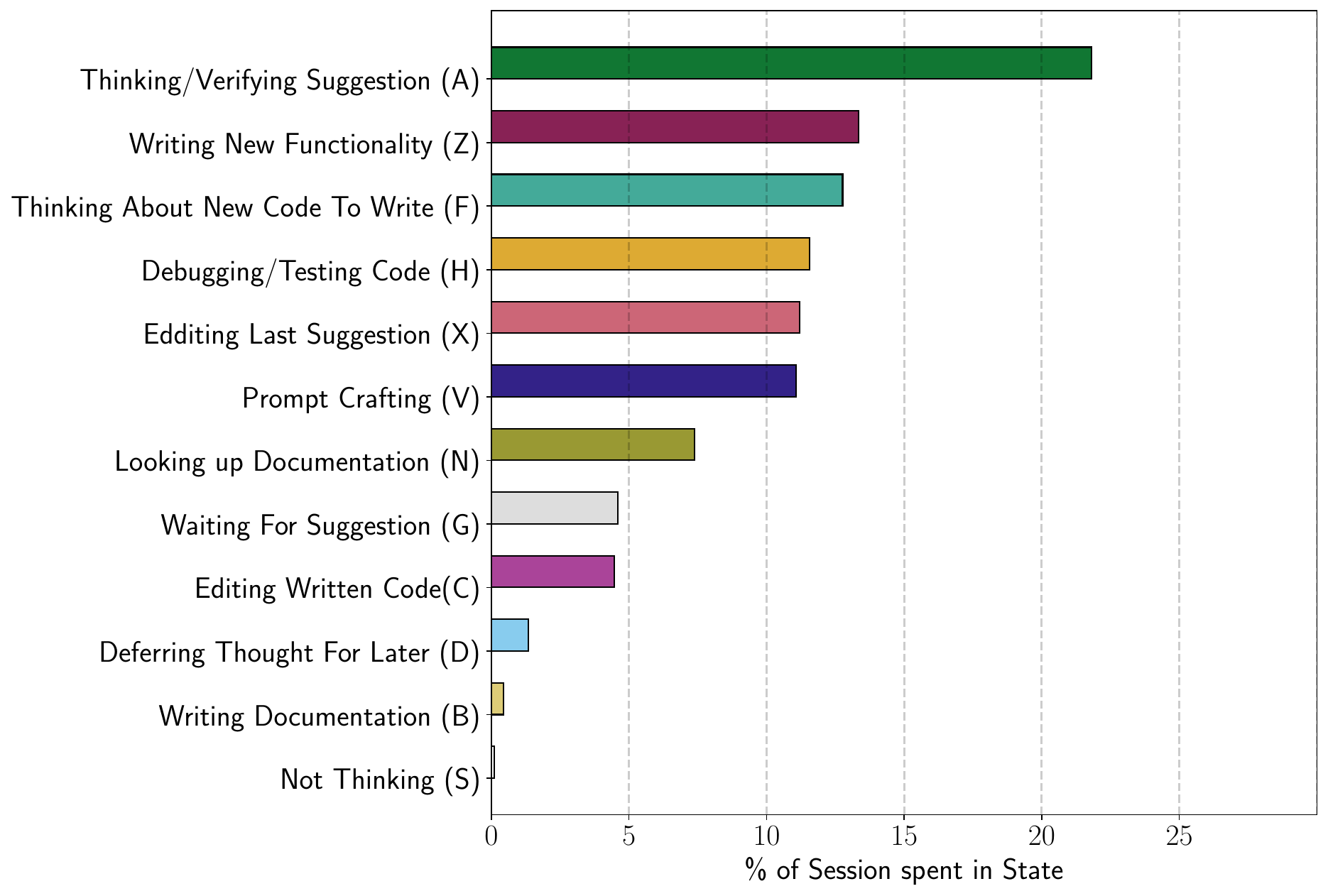}
    \caption{
   The percentage of total session time spent in each state during a coding session. On average, verifying \copilot suggestions occupies a large portion of session time.}
    \label{fig:histogram}
  \end{subfigure}
    \hfill
  \begin{subfigure}[b]{0.55\textwidth}
    \includegraphics[width=\textwidth]{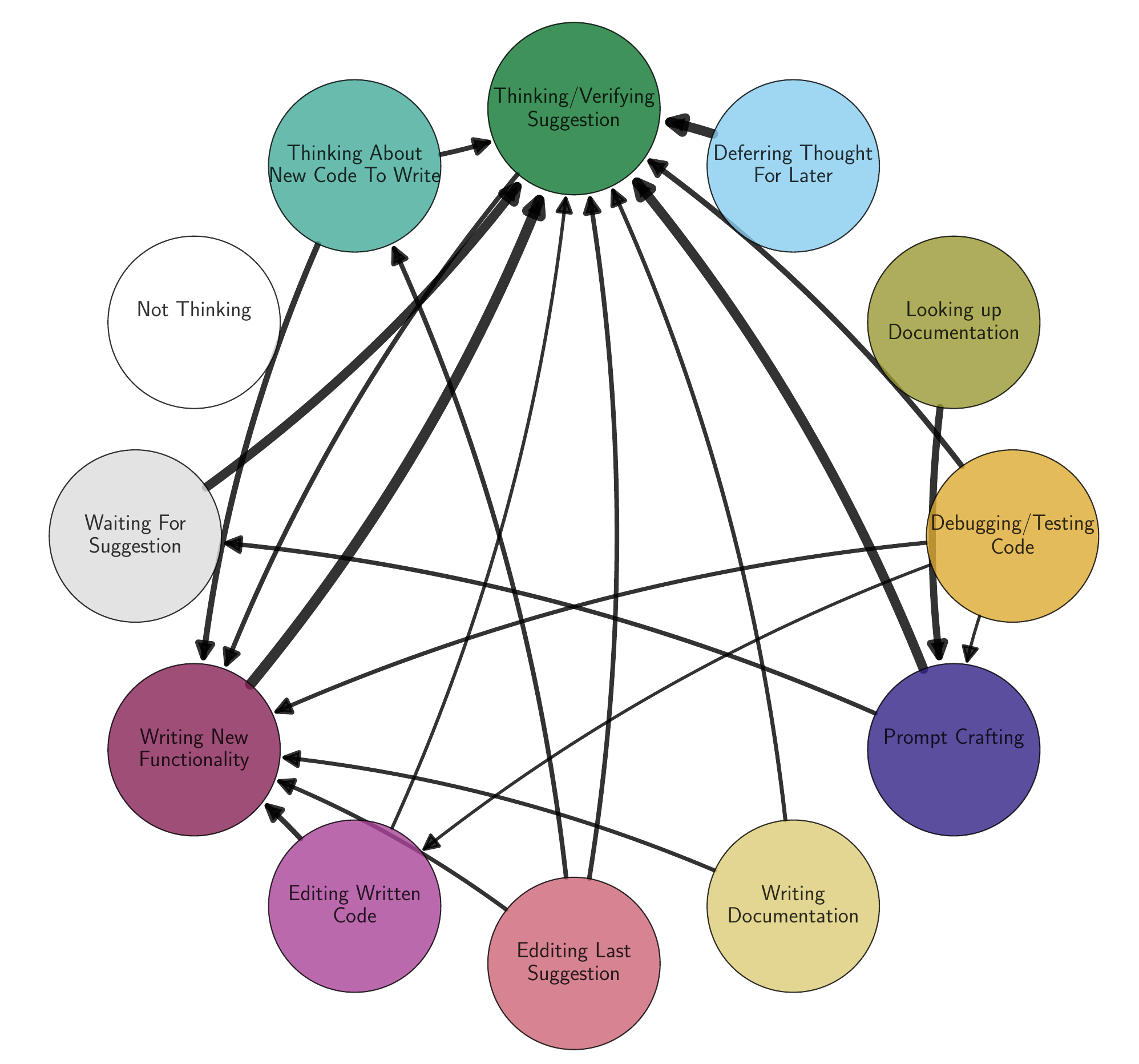}
    \caption{\cups diagram showing 12 CUPS states (nodes) and the transitions among the states (arcs). Transitions occur when a suggestion is shown, accepted, or rejected. We hide self-transitions and low-probability transitions for simplicity}
    \label{fig:cups_diagram}
  \end{subfigure}

  \caption{
  Visualization of \cups labels from our study as timelines, a histogram, and a state machine.
  }
  \label{fig:overall_cups}

\end{figure*}

\section{Understanding Programmer Behavior with \cups: Main Results} \label{sec:study_results}
The study in the previous section allows us to collect telemetry with {\cups} labels for each telemetry segment. We now analyze the collected data and highlight suggestions for 1) \textbf{metrics} to measure the programmer-{\coderec} interaction, 2) \textbf{design improvements} for the {\copilot} interface, and finally 3) \textbf{insights} into programmer behavior. Each subsection below presents a specific result or analysis which can be read independently. Code and Data is available at \footnote{\url{https://github.com/microsoft/coderec_programming_states}}.

\subsection{Aggregated Time Spent in Various CUPSs} \label{subsec:time_in_cups}

In Figure~\ref{fig:timelines}, we visualize the coding sessions of {\em individual} participants as {\em \cups timelines}, where each telemetry segment is labeled with its \cups label.
At first glance, \cups timelines show the richness in patterns of interaction with \copilot, as well as the variance in usage patterns across settings and people.
\cups timelines allow us to inspect individual behaviors and identify patterns, which we later aggregate to form general insights into user behavior.

Figure \ref{fig:histogram} shows the average time spent in each state as a percentage normalized to a user's session duration. 

\yellowbox{\textbf{Metric Suggestion}: Time spent in \cups states as a high-level diagnosis of the interaction}{
For example, time spent `Waiting For Suggestion' (4.2\%, $s_N=4.46$ ) measures the real impact of \textbf{latency}, and time spent `Editing Last Suggestion' provides feedback on the quality of suggestions.
}

We find that averaged across all users, the \textbf{ `verifying suggestion' state takes up the most time} at 22.4\% ($s_N=12.97$), it is the top state for 6 participants and in the top 3 states for 14 out of 21 participants taking up at least 10\% of session time for all but one participant.  Notably, this is a new programmer task introduced by \copilot. The second-lengthiest state is writing new functionality' 14.05\% ($s_N=8.36$), all but 6 participants spend more than 9\% of session time in this state.

More generally, the states that are specific to interaction with \copilot include: `Verifying Suggestions', `Deferring Thought for Later', `Waiting for Suggestion', `Prompt Crafting', and `Editing Suggestion'. {\bf We found that the total time participants spend in these states is 51.5 \% ($s_N=19.3$) of the average session duration.} In fact, half of the participants spend more than 47\% of their session in these {\copilot} states, and all participants spend more than 21\% of their time in these states.

\textbf{By Programmer Expertise and \copilot Experience.} We investigate if there are any differences in how programmers interacted with {\copilot} based on their programming expertise and their previous experience with {\copilot}.  First, we split participants based on whether they have professional programming experience of more than 6 years (10 out of 21) and who have less than 6 years (11 out of 21).  We notice the acceptance rate for those with substantial programming experience is 30.0\% $\pm$ 14.5 while for those without is 37.6\% $\pm$ $14.6$. Second, we split participants based on whether they had used  {\copilot} previously (10 out of 21) and those who had never used it before (11 out of 21). The acceptance rate for those who have previously used {\copilot} is 37.6 \% $\pm$ 15.3, and for those who have not, it is $29.3$ $\pm$ 13.7. Due to the limited number of participants, these results are not sufficient to determine the influence of programmer experience or \copilot experience on behavior. \revision{We also include in Appendix a breakdown of programmer behavior by task solved.}
\subsection{Patterns in Behavior as Transitions Between \cups States } \label{subsec:patterns}
To understand if there was a pattern in participant behavior, we modeled {\em transitions} between two states as a {\em state machine}. We refer to the state machine-based model of programmer behavior as a {\em \cups diagram}. In contrast to the timelines in Figure~\ref{fig:timelines}, which visualize state transitions with changes of colors, the \cups diagram Figure~\ref{fig:cups_diagram} explicitly visualizes transitions using directed edges, where the thickness of arrows is proportional to the likelihood of transition. For simplicity, Figure~\ref{fig:cups_diagram} only shows transitions with an average probability higher than 0.17 ($90\textsuperscript{th}$ quantile, selected for graph visibility).

The transitions in Figure~\ref{fig:cups_diagram} revealed many expected patterns. For example, one of the most likely transitions (excluding self-transitions from the diagram), `\textrm{Prompt  Crafting}  $\xrightarrow{0.54}$ \textrm{Verifying Suggestion}' showed that when programmers were engineering prompts, they were then likely to immediately transition to verifying the resultant suggestions (probability of 0.54). Likewise, Another probable transition was `\textrm{Deferring Thought}$\xrightarrow{0.54}$\textrm{Verifying Suggestion}', indicating that {\bf if a programmer previously deferred their thought for an accepted suggestion, they would, with high probability, return to verify that suggestion}. Stated differently: deference incurs verification dept, and this debt often ``catches up'' with
the programmer. Finally, the single-most probable transition, `$\textrm{Writing New Functionality}\xrightarrow{0.59}\textrm{Verifying Suggestion}$', echos the observation from the previous section, indicating that programmers often see suggestions while writing code (rather than prompt crafting), then spend time verifying it. If suggestions are unhelpful, they could easily be seen as interrupting the flow of writing.

The \cups diagram also revealed some unexpected transitions. Notably, the second-most probable transition from the `Prompt Crafting' state is `$\textrm{Prompt  Crafting}\xrightarrow{0.25}\textrm{Waiting for Suggestion}$'. This potentially reveals an unexpected and unnecessary delay and is a possible target for refinement (e.g., by reducing latency in \copilot). 
Importantly, each of these transitions occurs with a probability that is much higher than the lower bound/uniform baseline probability of transitioning to a random state in the \cups diagram (1/12=0.083). In fact, when we compute the entropy rate (a measure of randomness) of the resulting Markov Chain \cite{ekroot1993entropy} from the \cups diagram we obtain a rate of 2.24; if the transitions were completely random the rate would be 3.58, and if the transitions were deterministic then the rate is 0. 

\yellowbox{\textbf{Interface Design Suggestion}: Identifying current \cups state can help serve programmer needs}{
If we are able to know the current programmer \cups state during a coding session we can better serve the programmer, for example,
\begin{itemize}
    \item If the programmer is observed to have been deferring their thought on the last few suggestions, group successive \copilot suggestions and display them together. 
    \item If the programmer  is waiting for the suggestion, we can prioritize resources for them at that moment
    \item While a user is prompt crafting, \copilot suggestions are often ignored and may be distracting; however, after a user is done with their prompt, they may expect high-quality suggestions. We could suppress suggestions during prompt crafting, but after the prompt crafting process is done, display multiple suggestions to the user and encourage them to browse through them. 
\end{itemize}
Future work can, for example, realize these design suggestions by allowing \textbf{custom keyboard macros} for the programmer to signal their current \cups state, or a more automated approach by \textbf{predicting} their \cups state.
}

\begin{figure*}
   \begin{subfigure}{1.0\textwidth}
   \centering
    \includegraphics[height=8cm,width=\textwidth]{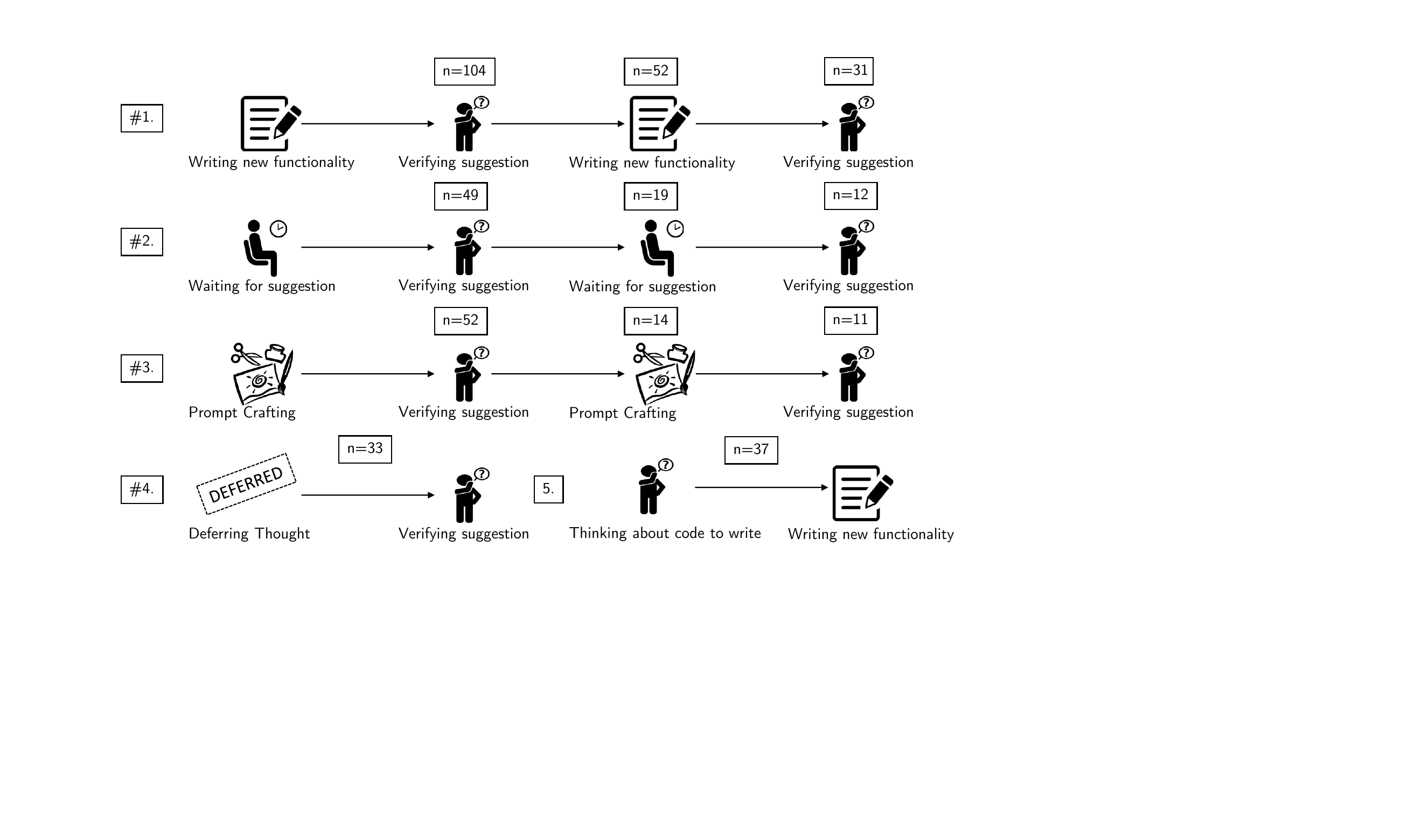}
    \caption{Common patterns of transitions between {\em distinct} states. 
     In individual participant timelines, the patterns visually appear as a change of color,
     but here we measure how often they appear across all participants (n=).}
    \label{fig:common_sequences_colorchange}
  \end{subfigure}
  \hfill
  \begin{subfigure}{1.0\textwidth}
     \centering

    \includegraphics[width=\textwidth]{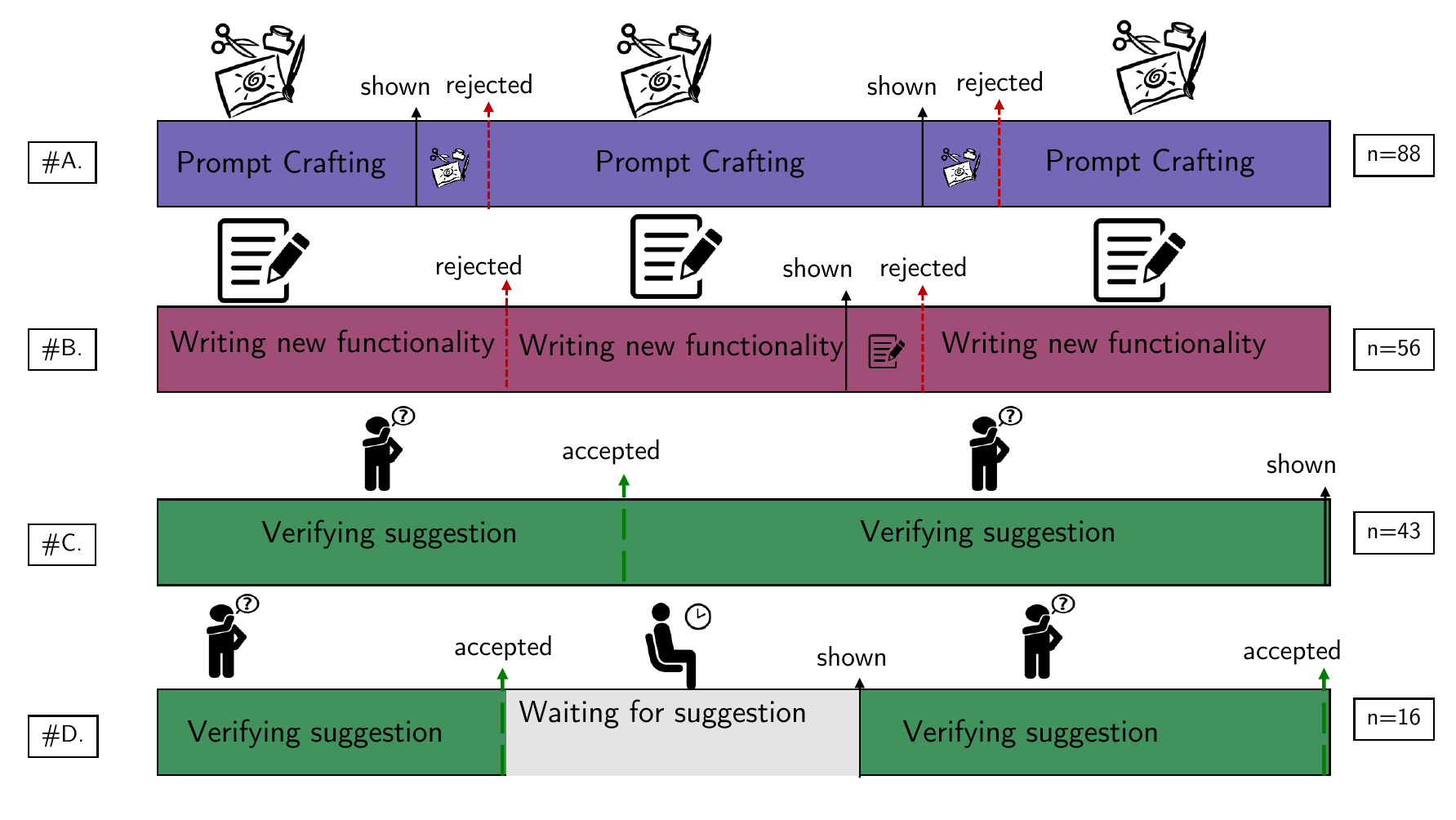}
    \caption{Common patterns of states and actions (including self transitions).
    Each pattern is extracted from user timelines and we count how often it appears in total (n=)}
    \label{fig:common_sequences_segmentchange}
  \end{subfigure}
  \hfill
  \caption{Myriad of \cups patterns observed in our study. }

\end{figure*}

We also investigated longer patterns in state transitions by searching for the most common sequence of states of varying lengths. We achieved this by searching over all possible segment n-grams and counting their occurrence over all sessions. We analyzed patterns in two ways:  in Figure \ref{fig:common_sequences_colorchange}, we merged consecutive segments that have the same state label into a single state (thus removing self-transitions), and in Figure \ref{fig:common_sequences_segmentchange} we looked at n-grams in the user timelines (including self-transitions) where we include both states and participants actions (shown, accepted and rejected). The most common pattern (\#1) in Figure \ref{fig:common_sequences_colorchange} was a cycle where programmers repeatedly wrote new code functionality and then spent time verifying shown suggestions, indicating a new mode for programmers to solve coding tasks. At the same time, when we look at pattern (\#B) in Figure \ref{fig:common_sequences_segmentchange}, which takes a closer look into when programmers are writing new functionality, we observe that they don't stop to verify suggestions and reject them as they continue to write. 
Other long patterns include \revision{(\#2)}  (also shown as pattern \#D ), where programmers repeatedly accepted successive \copilot suggestions after verifying each of them. Finally, we observe in \revision{(\#3)}  and (\#A)  programmers iterating on the prompt for \copilot until they obtain the suggestion they want. We elaborate more on this in the next subsection.

\subsection{Programmers Often Defer Thought About Suggestions}\label{subsec:defer_thought}

\begin{figure*}[ht]
    \centering
    \includegraphics[trim ={0 1cm 0 0} ,width=\textwidth]{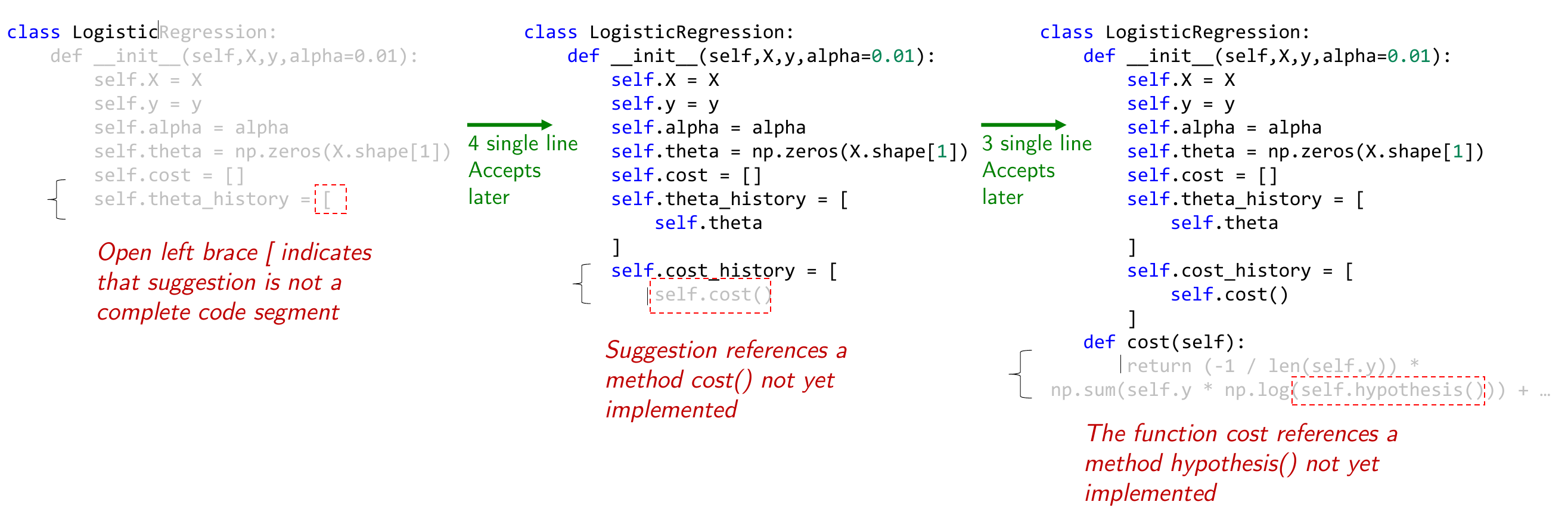}
    \caption{
    Illustration of a coding scenario with \copilot where the programmer may choose to defer verifying a suggestion (`Deferring Thought'). Here, \copilot suggests an
    implementation for the class {\tt Logistic Regression} 
    line-by-line (illustrated from left to right). And the programmer may need to defer verifying intermediate suggestion of {\tt self.cost} (middle screenshot) because the method that implemented it is suggested later (right screenshot).
    }
    \label{fig:defer_thought}

\end{figure*}

 An interesting {\cups} state is that of 'Deferring Thought About A Suggestion'. This is illustrated in Figure {\ref{fig:defer_thought}}, where programmers accept a suggestion or series of suggestions without sufficiently verifying them beforehand. This occurs either because programmers wish to see the suggestion with code highlighting, or because they want to see where {\copilot} suggestions leads to.
Figure~\ref{fig:histogram} shows that programmers do in fact, frequently defer thought-- we counted 61 states labeled as such.
What drives the programmer to defer their thought about a suggestion rather than immediately verifying it? We initially conjectured that the act of deferring may be partially explained by the length of the suggestions. So, we compared the number of characters and the number of lines for suggestions depending on the programmer's state. We find that there is no statistical difference according to a two-sample independent t-test $(t=-0.58, p=0.56)$\footnote{All p-values reported are corrected for multiple hypothesis testing with the Benjamin/Hochberg procedure with $\alpha=0.05$.} in the average number of characters between deferred thought and suggestions (75.81 compared to 69.06) that were verified previously. The same holds for the average number of lines. 

However, when we look at the likelihood of editing an accepted suggestion, we find that it is 0.18 if it was verified before, but it is 0.53 if it was deferred. This difference is significant according to a chi-square test $(\chi^2=29.2,p=0)$. In fact, the programmer CUPS state has a big effect on their future actions. In Table \ref{tab:prob_Accept_by_state}, we show the probability of the programmer accepting a suggestion given the \cups state the programmer was in while the suggestion is being shown. We also show the probability of the programmer accepting a suggestion as a function of the \cups state the programmer was in just before the suggestion was displayed. We observe there is a big variation in the suggestion acceptance rate by the \cups state. For example, if the programmer was in the "Deferring Thought For Later" state, the probability of acceptance is 0.98 $\pm$ 0.02 compared to when a programmer is thinking about new code to write, where the probability is 0.12 $\pm$ 0.04. Note that the average probability of accepting a suggestion was 0.34.

{\em What are the programmers doing before they accept a suggestion?} 
We found that the average probability of accepting a suggestion was 0.34. However, we observed that when the programmer was verifying a suggestion their likelihood of accepting was 0.70. In contrast, 
if the programmer was thinking about new code to write, the probability dropped to 0.20. This difference was statistically significant  according to Pearson's chi-squared test $(\chi^2=12.25,p=0)$. 
Conversely, when programmers are engineering prompts, the likelihood of accepting a suggestion drops to 0.16. One reason for this might be that programmers want to write the prompt on their own without suggestions, and \copilot interrupts them.  We show the full results in the Appendix for the other states.

\begin{table}[ht]
\caption{ We compute the percentage of suggestions accepted given the programmer was in the \cups state while the suggestion is being shown (\% Ss accepted while shown). We compute the percentage of suggestions accepted given the programmer was in the \cups state before the suggestion is shown, the state just before the one where the suggestion is shown (\% Ss accepted before S is shown). We compute the standard error for the acceptance rate (\%).}
\begin{tabular}{p{7cm}p{4cm}p{4cm}}
    \toprule
\textbf{State} & \textbf{\% Ss accepted while shown}  & \textbf{\% Ss accepted before S is shown}  \\ 
\midrule
Thinking/Verifying Suggestion & 0.80 $\pm$ 0.02 & 0.56 $\pm$ 0.04 \\ 
Prompt Crafting & 0.11 $\pm$ 0.02 & 0.22 $\pm$0.03\\ 
Looking up Documentation & 0.00 $\pm$ 0.00 & 0.29 $\pm$ 0.17\\ 
Writing New Functionality & 0.07 $\pm$ 0.02 & 0.31 $\pm$ 0.03\\ 
Thinking About New Code To Write & 0.12 $\pm$ 0.04 & 0.27 $\pm$ 0.04\\ 
Editing Last Suggestion & 0.03 $\pm$ 0.03 & 0.23 $\pm$ 0.05\\ 
Waiting For Suggestion & 0.10 $\pm$ 0.05 & 0.58 $\pm$ 0.06\\ 
Editing Written Code& 0.07 $\pm$ 0.04 & 0.17 $\pm$ 0.07\\ 
Writing Documentation& 0.40 $\pm$ 0.22 & 0.33 $\pm$ 0.19\\ 
Debugging/Testing Code& 0.23 $\pm$ 0.07 & 0.26 $\pm$ 0.06\\ 
Deferring Thought For Later& 0.98 $\pm$ 0.02 & 1.0 $\pm$ 0.0\\ 
  \bottomrule
\end{tabular}
    \label{tab:prob_Accept_by_state}

\end{table}

\subsection{\cups Attributes Significantly More Time Verifying Suggestions than Simpler Metrics}\label{subsec:time_adjustment}

\begin{figure*}[t]
    \centering
    \includegraphics[trim={0 2cm 0 0},width=0.8\textwidth]{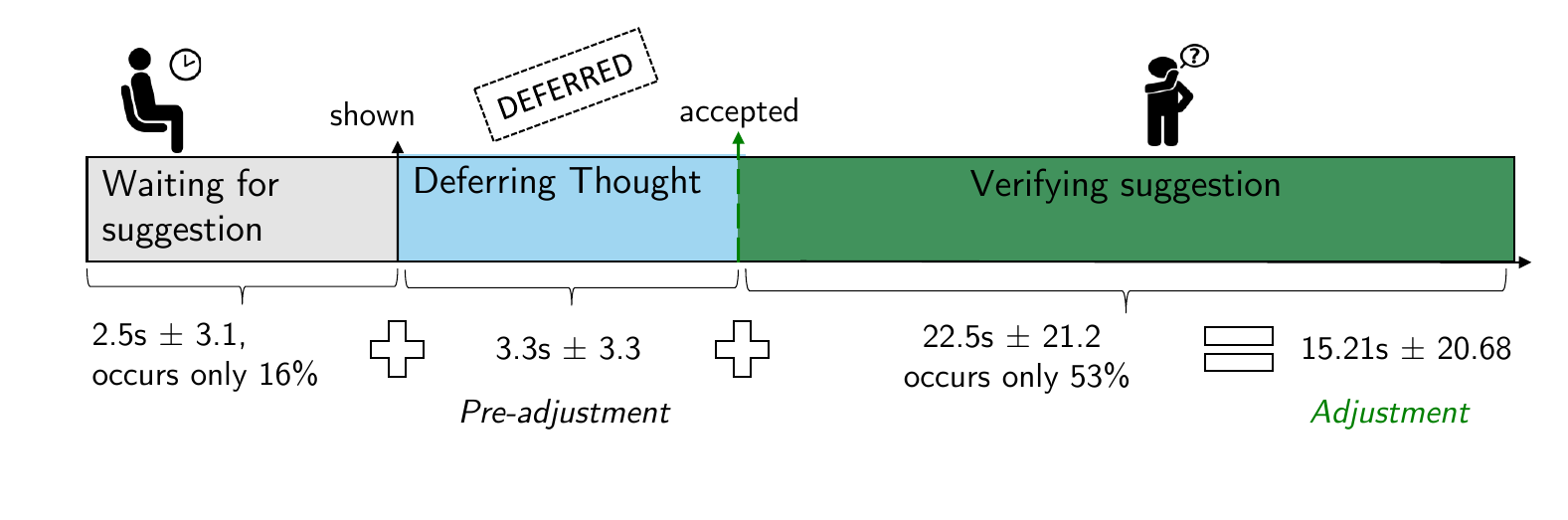}
    \caption{Illustration of one of the adjustments required for measuring the total time a programmer spends to verify a suggestion. Here, when a programmer defers thought for a suggestion, they spend time verifying it after accepting it and may also have to wait beforehand for the suggestion to be shown. }
    \label{fig:adjustment_time}

\end{figure*}

We observed that programmers continued verifying the suggestions after they accepted them.
This happens by definition for 'deferred thought' states before accepting suggestions, but we find it also happens when programmers verify the suggestion before accepting it and this leads to a significant increase in the total time verifying suggestions. First, when participants defer their thought about a suggestion they accepted, 53.2\% of the time they verify the suggestion immediately afterward. When we adjust for the post-hoc time spent verifying, we compute a mean time of $15.21$ \sdev{20.68} seconds of verification and a median time of $6.48$s. This is nearly a five-times increase in average time and a three-time increase in median time for the pre-adjustment scores of $3.25$ \sdev{3.33} mean and $1.99$ median time. These results are illustrated in Figure \ref{fig:adjustment_time} and is a statistically significant increase according to a two-sample paired t-test $(t=-4.88, p=1.33 \cdot 10^{-5})$. This phenomenon also occurs when programmers are in a 'Thinking/Verifying Suggestion' state before accepting a suggestion where 19\% of the time they posthoc verify the suggestion which increases total verification time from $3.96$ \sdev{8.63} to $7.03$ \sdev{14.43} on average which is statistically significant $(t=-4.17,p=5e-5)$. On the other hand, programmers often have to wait for suggestions to show up due to either latency or \copilot not kicking in to provide a suggestion. If we sum the time between when a suggestion is shown and the programmer accepts or rejects this in addition to the time they spend waiting for the suggestion (this is indicated in the state 'Waiting for suggestion'), then we get an increase from $6.11$s \sdev{15.52} to $6.51$s \sdev{15.61} which is minor on average but adds 2.5 seconds of delay when programmers have to explicitly wait for suggestions.

\yellowbox{\textbf{Metric Suggestion}: Adjust verification time metrics and acceptance rates to include suggestions that are verified after acceptance}{
The previous analysis showed that the time to accept a suggestion cannot be simply measured as the time spent from the instance a suggestion is shown until a suggestion is accepted-- this misses the time programmers spend verifying a suggestion after acceptance. Similarly, since deferring thought is a frequent behavior observed, it leads to an inflation of acceptance rates. We recommend using measures such as the fraction of suggestions accepted that survive in the codebase after a certain time period (e.g. 10 minutes).
}

\subsection{Insights About Prompt Crafting}\label{subsec:prompt_craft}

\paragraph{Insights about Prompt Crafting.} We take a closer look into how participants craft prompts to obtain \copilot suggestions. \emph{Our first insight is that programmers consistently ignore suggestions while prompt crafting. Among 234 suggestions that were shown while participants were actively prompt crafting, defined as a suggestion where a programmer was prompt crafting while the suggestion was being displayed, only 10.7\% were accepted.} We hypothesize this behavior could be due to programmers wanting to craft the prompt in their own language rather than relying on \copilot to help them prompt craft. This also indicates that \copilot is unnecessarily interrupting participants' prompt crafting attempts. 

However, programmers often iterate on their prompts until they obtain the suggestion they desire and often do not abandon prompt crafting without accepting a suggestion. We define a prompt crafting attempt as a segment of the coding session that starts from when the programmer first enters the \cups "prompt crafting" state and lasts until the programmer enters a non-\copilot centric state \footnote{The non-\copilot centric states are: 'Writing New Functionality,' 'Editing Written Code,''Thinking About New Code To Write,''Debugging\/Testing Code,' 'Looking up Documentation,' 'Writing Documentation.'}. We count 59 such prompt crafting attempts wherein 81.3\% of them a suggestion is accepted.

Prompt crafting is often an iterative process, where the programmer writes an initial prompt, observes the resulting suggestion, then iterates on the prompt by adding additional information about the desired code or by rewording the prompt.  For example, P5 wanted to retrieve the index of the maximum element in a correlation matrix and wrote this initial prompt and got the suggestion:
\begin{verbatim}
# print the indices of the max value excluding 1 in corr
maxval = np.amax(corr, axis=1) # Copilot suggestion
\end{verbatim}
This code snipped returns the value of the maximum value rather than the index, so it was not accepted by the participant. They then re-wrote the prompt to be:
\begin{verbatim}
# print  the two features  most correlated
# Copilot suggestion
maxcor = np.where(corr == np.amax(corr)) 
\end{verbatim}
and accepted the above suggestion. 

Finally, we observe that there are three main ways participants craft prompts:

\noindent 1) through writing a single line comment with natural language instructions, although the comment may resemble pseudo-code \cite{jiang2022discovering}, an example:
\begin{verbatim}
# impute missing values in X_train as average of column 
# where missinggn value is -1
\end{verbatim}
2) through writing a docstring for the function:
\begin{verbatim}
def distance(self, query):
    ''' 
    query: single numpy arrray
    return:  l2 distances from query to the vectors
    '''
\end{verbatim}
and finally, 3) through writing function signatures (or variable names) e.g., writing "def add\_time" then pausing to wait for a suggestion. Often, programmers combine the three prompt crafting strategies to get better code suggestions.

\subsection{Post-Study Survey Answers}\label{subsec:survey_insights}

After completing the study, participants were asked to complete a survey based on the productivity survey in \cite{ziegler2022productivity}, which focuses on the SPACE framework of programmer productivity \cite{forsgren2021space}. We also included a free-form text box at the end of the survey where participants can add any additional thoughts about their experience using \copilot for the task assigned. The full results of the survey can be found in the Appendix.

 We found that 6/21 participants agreed or strongly agreed with the statement that they were concerned about the quality of their code when using \copilot. Participant \#9 noted, "I worry that bugs can sneak-in and go unnoticed, especially in weakly-dynamically typed languages" and Participant \#19 noted that "My main concern with Copilot is whether it is teaching me to do things the wrong (or old) way (e.g. showing me a Python 3.6 way instead of a Python 3.10 way and so on)". On the other hand, 14/21 participants agreed that using \copilot helped them stay in flow and spend less time searching for information.  Participant \#3 noted that "Collaborating with Copilot felt like I was googling what I wanted to do except instead of going through several stack overflow links that Google would show me, the code just appeared inline saving me time and keeping my flow of coding" and Participant \#6 "Going into the exercise I genuinely thought there would be a point when I pull up stack overflow. Because that's the kind of tiny stuff you sometimes need to search for. With copilot, it really reduced my worry of doing so." Finally, 17/21 participants agreed with the statement that by using \copilot, they completed the task faster, and 16/21 participants agreed that they were more productive using \copilot. These survey responses highlight the costs and benefits of writing code with \copilot and reinforcing existing results in \cite{ziegler2022productivity}.

\section{Limitations, Future Work and Conclusion}\label{sec:discussion}

\subsection{Limitations}

The observations from our study are limited by several decisions that we made. 
First, our participants solved time-limited coding tasks that were provided by us instead of real tasks they may perform in the \revision{real world. Furthermore, the selection of tasks was limited and did not cover all tasks programmers might perform.} \revision{We mostly conducted experiments with Python with only two participants using C++ and JavaScript} when \copilot is capable of completing suggestions for myriads of other languages. We also made an assumption about the granularity of telemetry where each segment at most contained one state when, in a more general setting, programmers may perform multiple activities within a single segment. We also did not capture longer-term costs of interacting, e.g., from accepting code with security vulnerabilities or longer horizon costs. To this end, security vulnerabilities and possible overreliance issues  \cite{pearce2022asleep,asare2022github,pearce2021can}, are important areas of research that we do not address in this paper.

\subsection{Future Work}

We only investigated a limited number of programmer behaviors using the \cups timelines and diagrams. There are many other aspects future work could investigate.

\paragraph{Predicting CUPS states.} To enable our insights derived in Section~\ref{sec:study_results}, we need to be able to identify the current programmer's CUPS state. An avenue towards that is building predictive models using labeled telemetry data that is collected from our user study. Ideally, we can leverage this labeled data to further label telemetry data from other coding sessions or other participants so that we can perform such analyses more broadly.
Specifically, the input to such a model would be the current session context, for example, whether the programmer accepted the last suggestion, the current suggestion being surfaced, and the current prompt. We can leverage supervised learning methods to build such a model from collected data. 
Such models would need to run in real-time during programming and predict at each instance of time the current user CUPS state. 
This would enable the design suggestions proposed to serve to compute various metrics proposed. For example, if the model predicts that the programmer is deferring thought about a suggestion, we can group suggestions together to display them to the programmer. 
In the Appendix, we built small predictive models of programmers \cups state using labeled study data. However, the current amount of labeled data is not sufficient to build highly accurate models. There are multiple avenues to improve the performance of these models: 1) simply collecting a larger amount of labeled data which would be expensive, 2) using methods from semi-supervised learning that leverage unlabeled telemetry to increase sample efficiency \cite{van2020survey}, and 3) collecting data beyond what is captured from telemetry such as video footage of the programmer screen (e.g. cursor movement) to be able to better predict with the same amount of data. 

\paragraph{Assessing Individual Differences}
There is an opportunity to apply the \cups diagram to compare different user groups and compare how individuals differ from an average user. Does the nature of inefficiencies differ between user groups? Can we personalize interventions? Finally, we could also compare how the \cups diagram evolves over time for the same set of users.

\paragraph{Effect of Conditions and Tasks on Behavior}
We only studied the behavior of programmers with the current version of \copilot.
Future work could study how behavior differs with different versions of \copilot -- especially when versions use different models. In the extreme, we could study behavior when \copilot is turned off.
The latter could help assess the {\em counterfactual} cost of completing the task without AI assistance and help establish whether and where \copilot suggestions add net value for programmers.
 For example, maybe the system did not add enough value because the programmer kept getting into prompt crafting rabbit holes instead of moving on and completing the functions manually or with the assistance of web search.
 
 Likewise, if developers create a faster version of \copilot with less latency, the \cups diagram could be used to establish whether it leads to reductions in time spent in the "Waiting for Suggestion" state.

\paragraph{Informing New Metrics}
Since programmers' value may be multi-dimensional, how can we go beyond code correctness and measure added value for users? If \copilot improves productivity, which aspects were improved? Conversely, if it didn't, where are the efficiencies? One option is to conduct a new study where we compare the \cups diagram with \copilot assistance with a counterfactual condition where the programmers don't have access to \copilot. And use the two diagrams to determine where the system adds value or could have added value. For example, the analysis might reveal \revision{that some code snippets}  are too hard for programmers to complete by themselves but much faster with \copilot because the cost of double-checking and editing \revision{the} suggestion is much less than the cost of spending effort on it by themselves.
Conversely, the analysis might reveal that a new intervention for helping engineer prompts greatly reduced people's times in ``Prompt Crafting''.

Another option is to design offline metrics based on these insights that developers can use during the model selection and training phase. For example, given that programmers spent a large fraction of the time verifying suggestions, offline metrics that can estimate this (e.g., based on code length and complexity) may be useful indicators of which models developers should select for deployment. Future work will aim to test the effectiveness of these design suggestions as well.

\paragraph{Beyond Programming.} We also hope our methodology is applied to study other forms of AI assistants that are rapidly being deployed. For example, one can make an analogous \cups taxonomy for writing assistants for creative writers or lawyers. 

\subsection{Conclusion}

We developed and proposed a taxonomy of common programmer activities (\cups) and combined it with real-time telemetry data to profile the interaction.
At present, \cups contains 12 mutually unique activities that programmers perform between consecutive \copilot actions (e.g., such as accepting, rejecting, and viewing suggestions).
We gathered real-world instance data of \cups by conducting a user study with 21 programmers within our organization, where they solved coding tasks with \copilot and retrospectively labeled \cups for their coding session.
We collected over 3137 instances of \cups and analyzed them to generate \cups timelines that show individual behavior and \cups diagrams that show aggregate insights into the behavior of our participants. We also studied the time spent in these states, patterns in user behavior, and better estimates of the cost (in terms of time) of interacting with \copilot.

Our studies with \cups labels revealed that when solving a coding task with \copilot, programmers 
may spend a large fraction of total session time (34.3\%) on just double-checking and editing \copilot suggestions, and spend more than half of the task time on \copilot related activities, together indicating that introducing \copilot into an IDE can significantly change user behavior.
We proposed new metrics to measure the interaction by computing the time spent in each {\cups} state and modification to existing time and acceptance metrics by accounting for suggestions that get verified only after they get accepted. We proposed a new interface design suggestion: if we allow programmers to signal their current state, then we can better serve their needs, for example, by reducing latency if they are waiting for a suggestion.  

\section*{Acknowledgments}
HM partly conducted this work during an internship at Microsoft Research (MSR). We acknowledge valuable feedback from colleagues across MSR and GitHub including Saleema Amershi, Victor Dibia, Forough Poursabzi, Andrew Rice, Eirini Kalliamvakou, and Edward Aftandilian.

\bibliography{ref}
\appendix
\newpage

\section{Details User Study}
\subsection{Interfaces}

\begin{figure}[H]
    \centering
    \includegraphics[width=\textwidth]{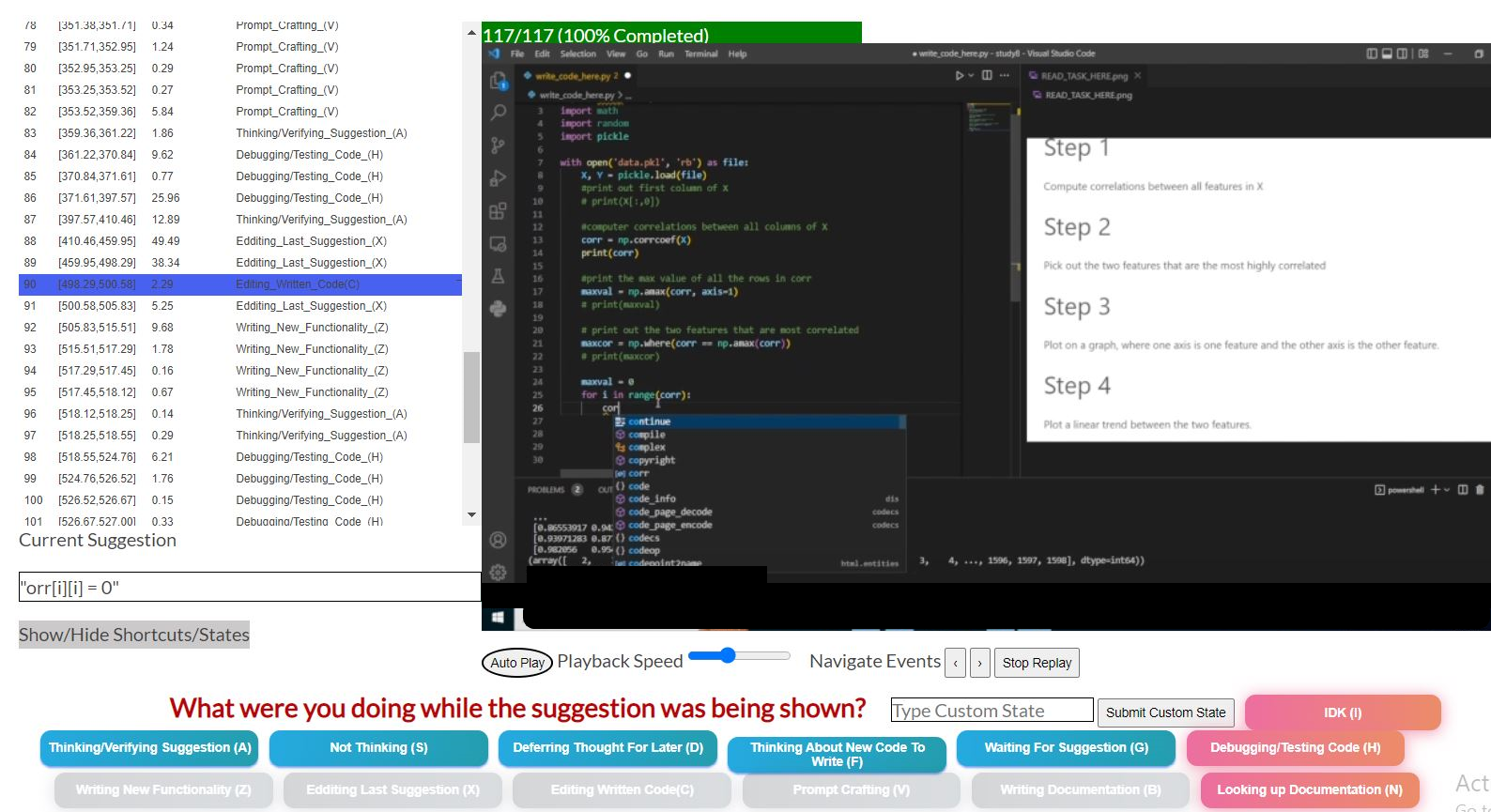}
    \caption{Screenshot of Labeling Tool represented in Figure \ref{fig:interface}}
    \label{fig:interface_screenshot}
\end{figure}

\begin{figure}[H]
    \centering
    \includegraphics[width=\textwidth]{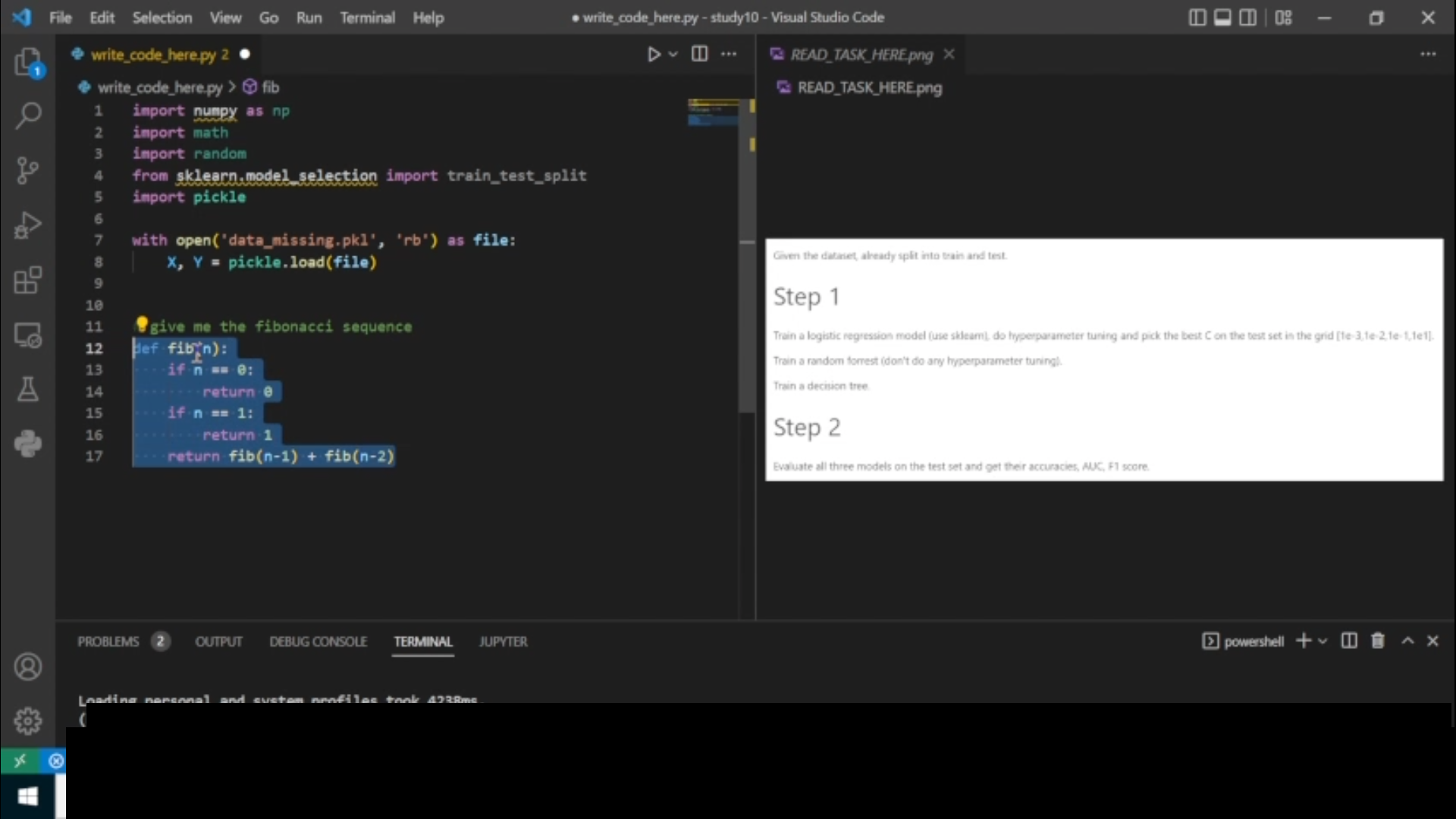}
    \caption{Screenshot of Virtual Machine interface with VS Code}
    \label{fig:vm_screenshot}
\end{figure}

\subsection{Task Instructions}

The tasks are shown to participants as image files to deter copying of the instructions as a prompt.

\begin{figure}[H]
    \centering
    \includegraphics[width=\textwidth]{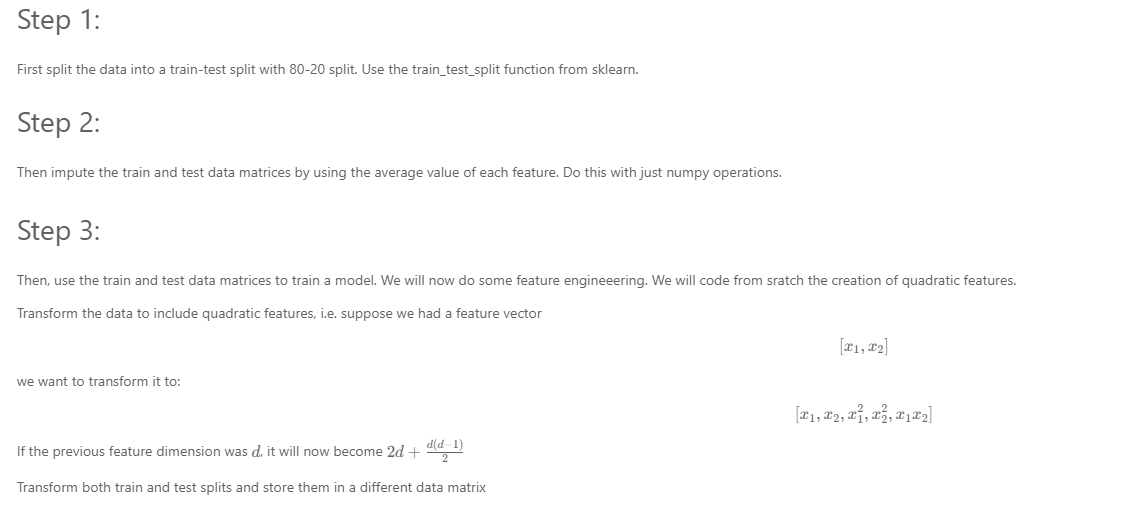}
    \caption{Data Manipulation Task.}
    \label{fig:task4}
\end{figure}

\begin{figure}[H]
    \centering
    \includegraphics[width=\textwidth]{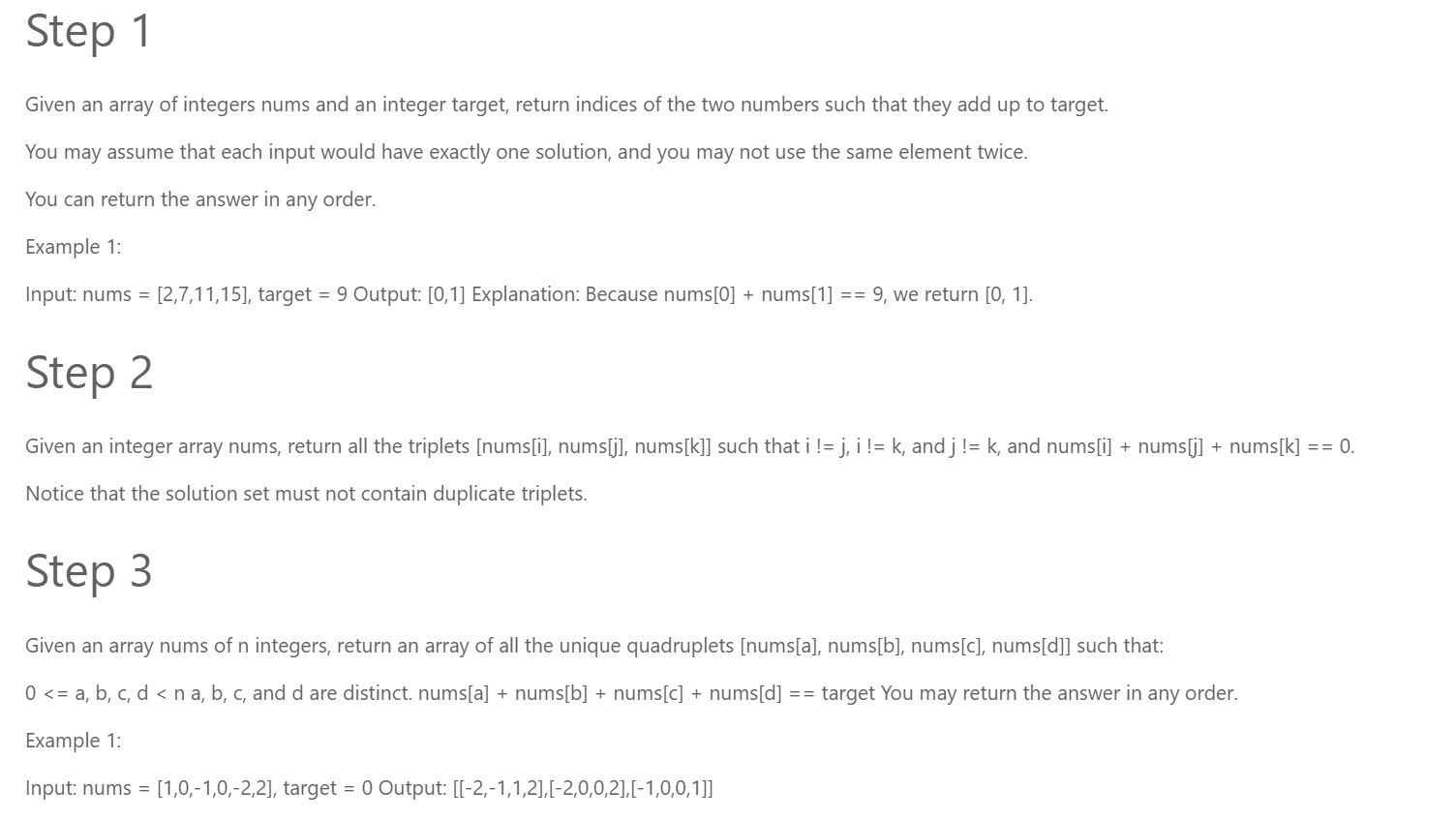}
    \caption{Algorithmic Problem Task.}
    \label{fig:task7}
\end{figure}

\begin{figure}[H]
    \centering
    \includegraphics[width=\textwidth]{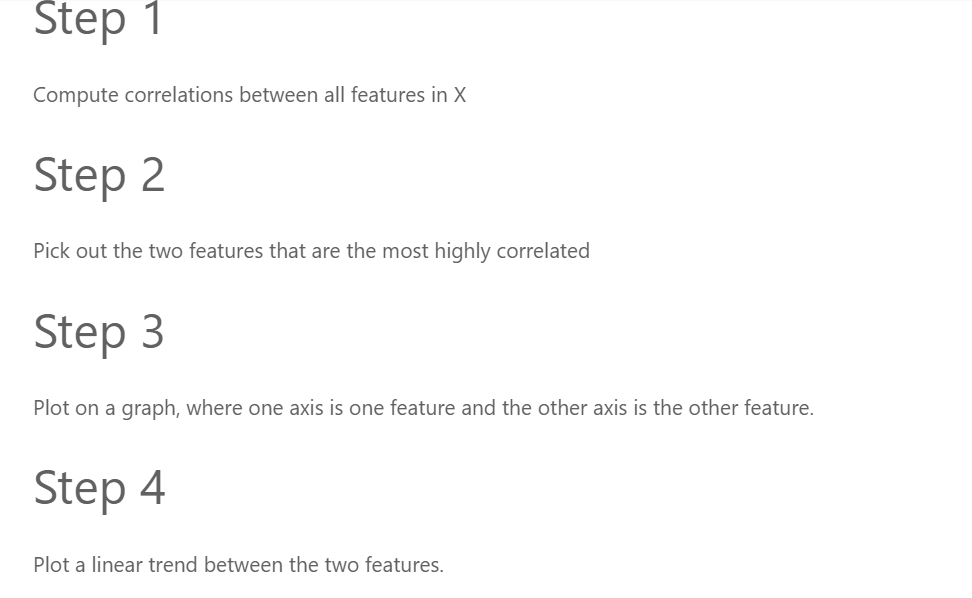}
    \caption{Data Analysis Task.}
    \label{fig:task8}
\end{figure}

\begin{figure}[H]
    \centering
    \includegraphics[width=\textwidth]{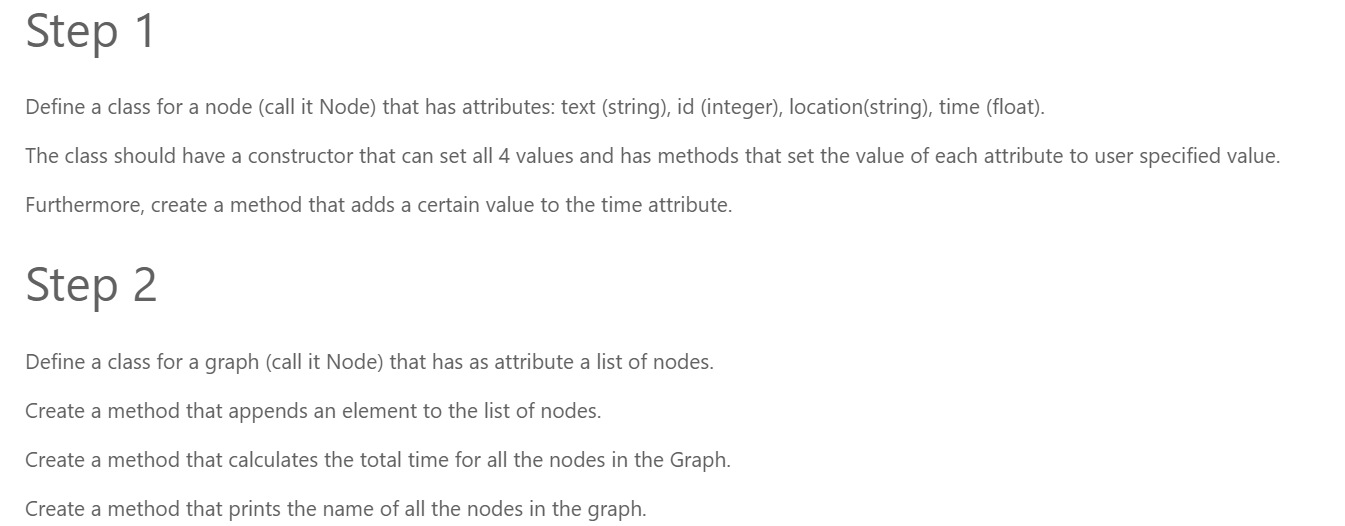}
    \caption{Classes and Boilerplate Code Task.}
    \label{fig:task9}
\end{figure}

\begin{figure}[H]
    \centering
    \includegraphics[width=\textwidth]{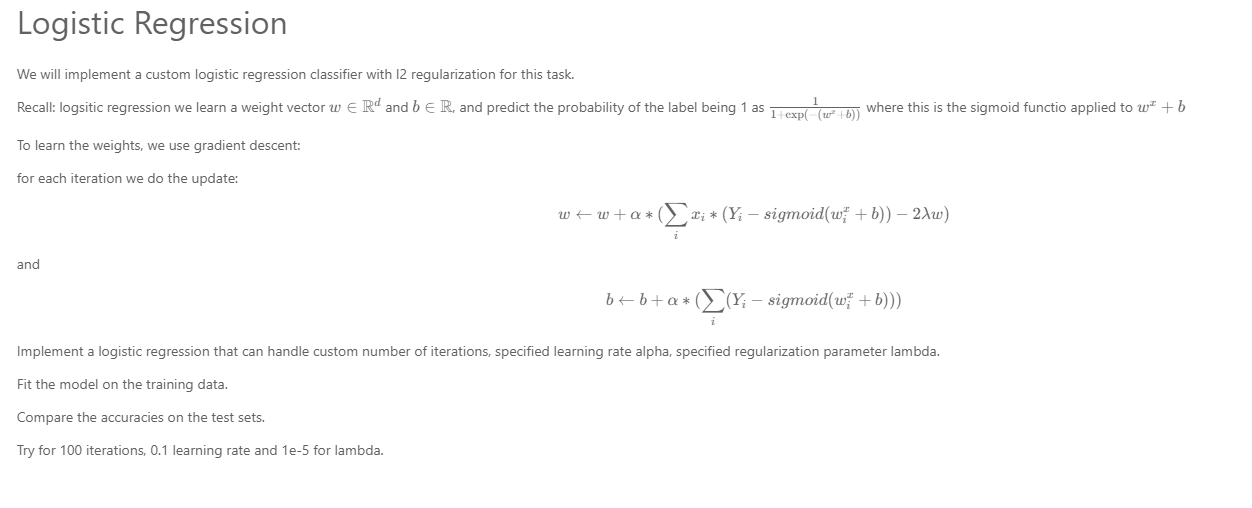}
    \caption{Logistic Regression Task}
    \label{fig:task16}
\end{figure}

\begin{figure}[H]
    \centering
    \includegraphics[width=\textwidth]{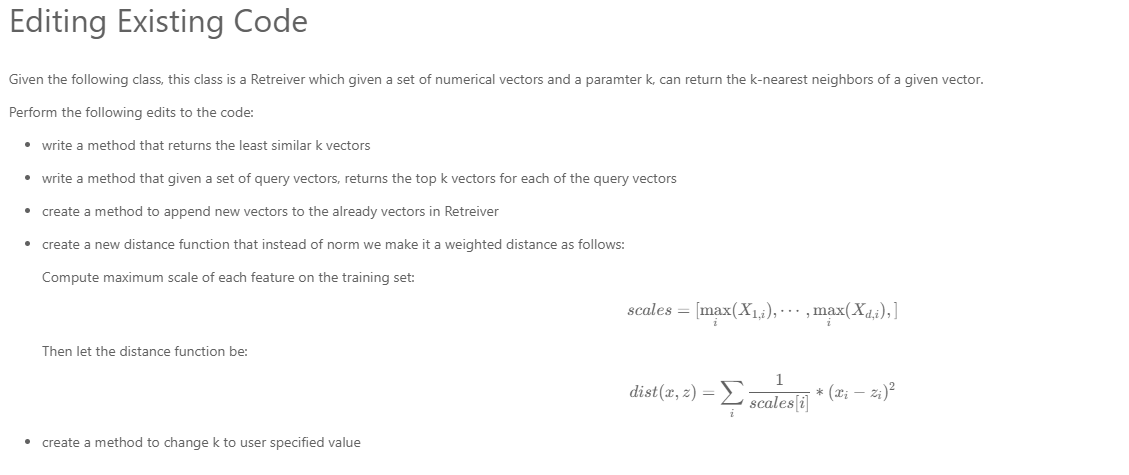}
    \caption{ Editing Code Task}
    \label{fig:task17}
\end{figure}

\begin{figure}[H]
    \centering
    \includegraphics[width=\textwidth]{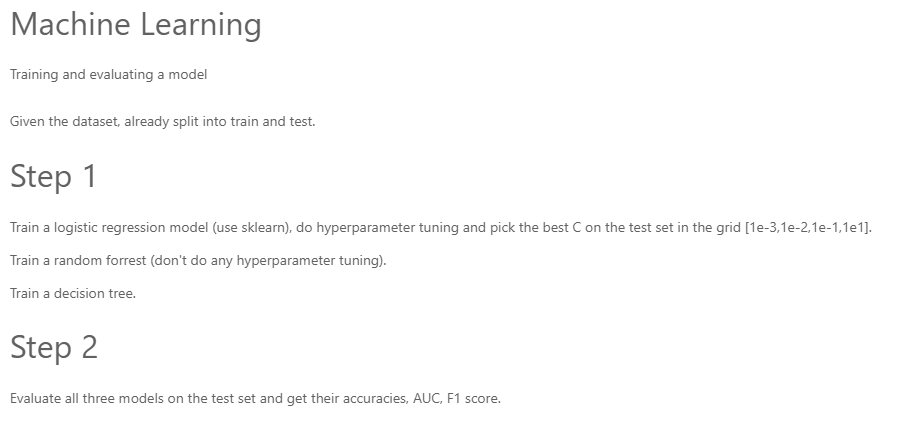}
    \caption{Machine Learning Task}
    \label{fig:task18}
\end{figure}

\begin{figure}[H]
    \centering
    \includegraphics[width=\textwidth]{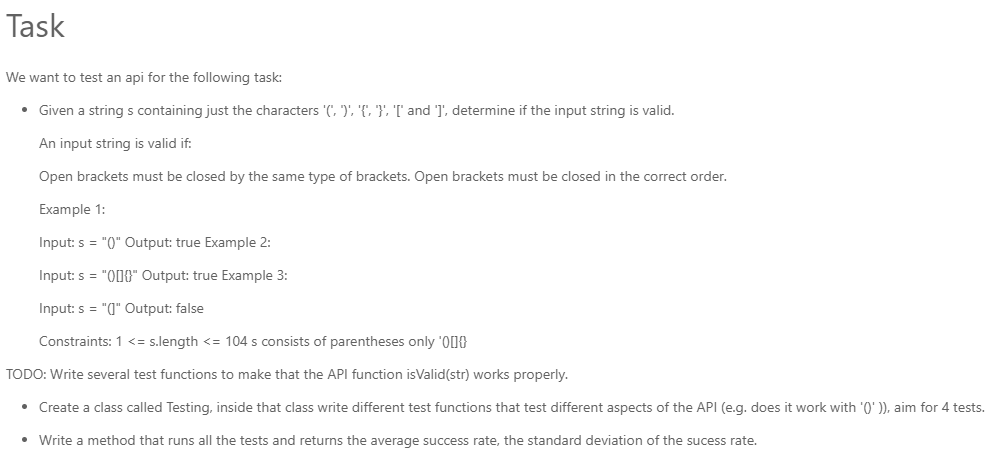}
    \caption{   Writing Tests Task}
    \label{fig:task19}
\end{figure}

\subsection{Survey Questions Results}

\begin{figure}[H]
    \centering
    \includegraphics[ height=18cm, width=14.5cm]{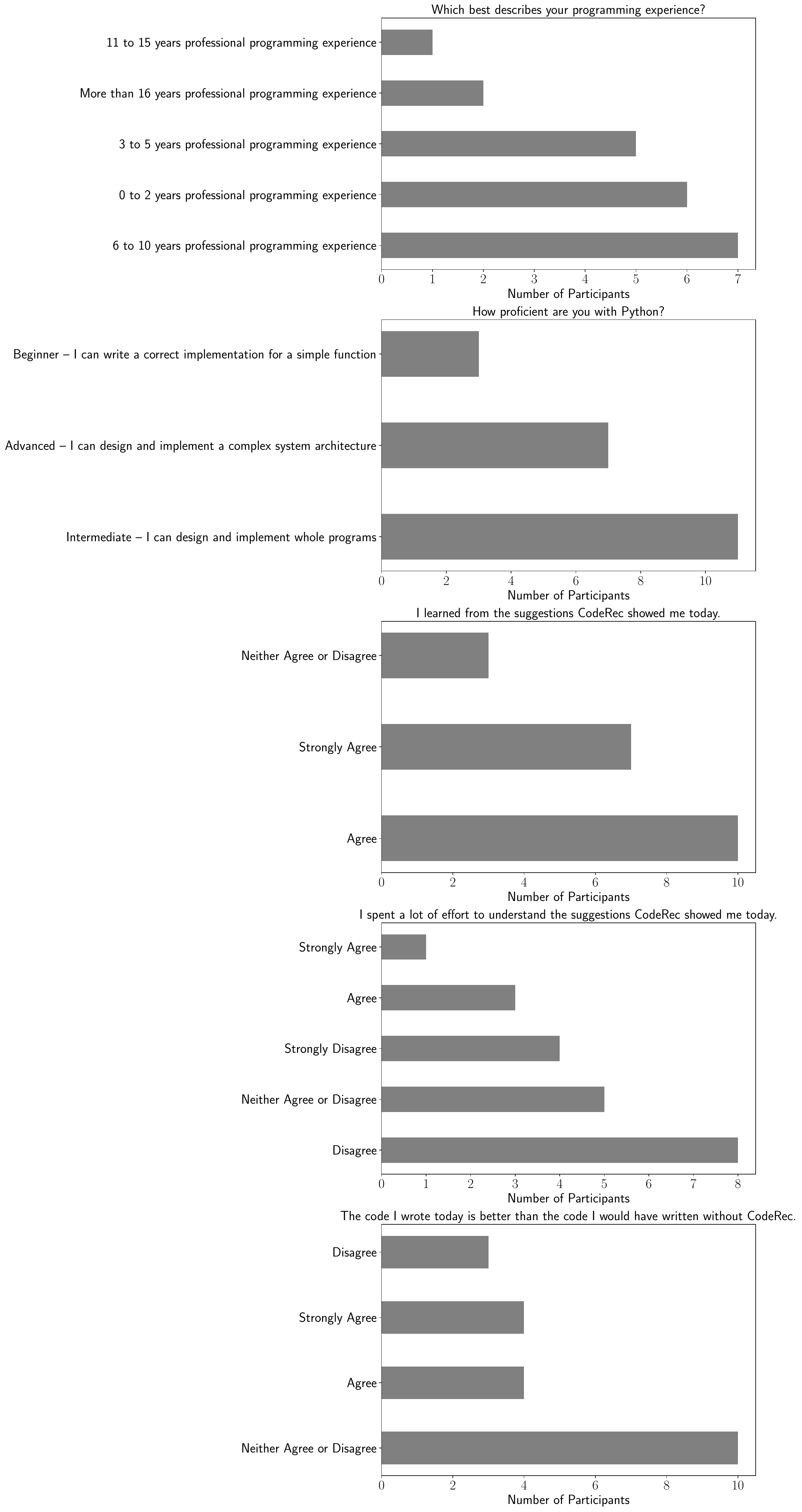}
    \caption{User Study Survey results (1)}
    \label{fig:survey1}
\end{figure}

\begin{figure}[H]
    \centering
    \includegraphics[ height=18cm, width=14.5cm]{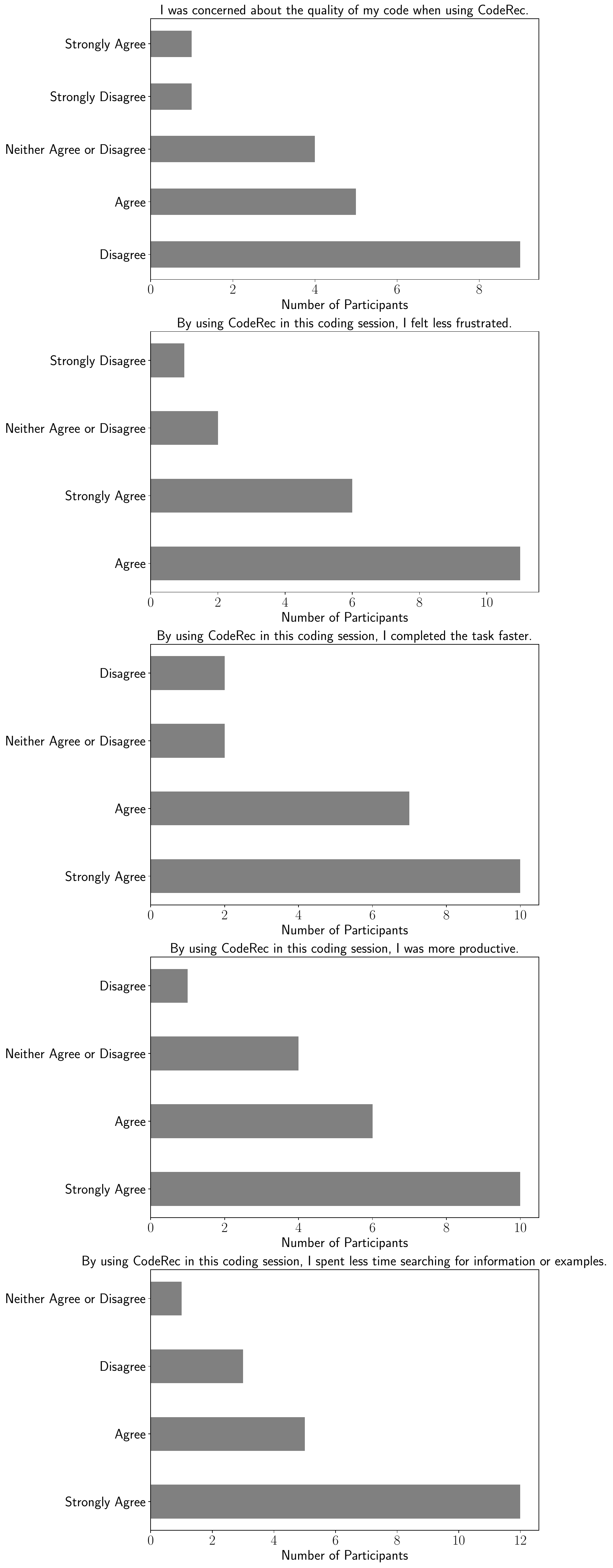}
    \caption{User Study Survey results (2)}
    \label{fig:survey2}
\end{figure}

\begin{figure}[H]
    \centering
    \includegraphics[ height=18cm, width=14.5cm]{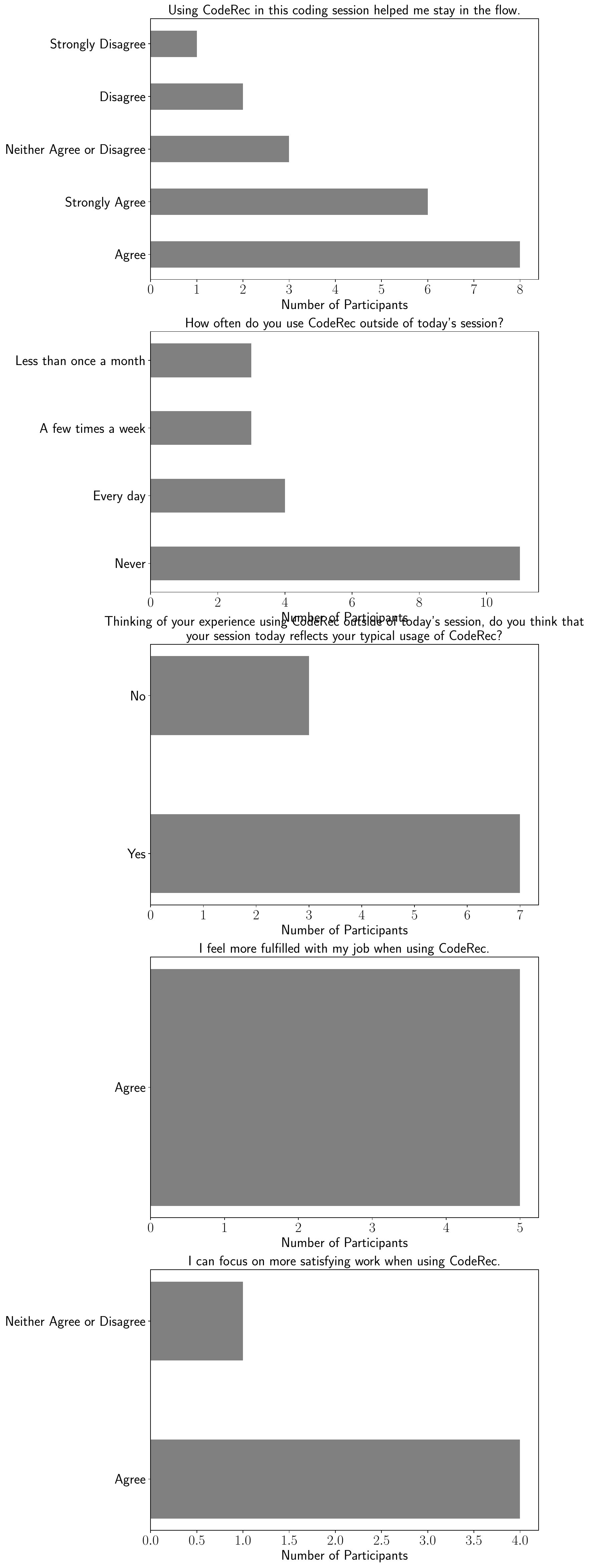}
    \caption{User Study Survey results (3)}
    \label{fig:survey3}
\end{figure}

\begin{figure}[H]
    \centering
    \includegraphics[ height=18cm, width=14.5cm]{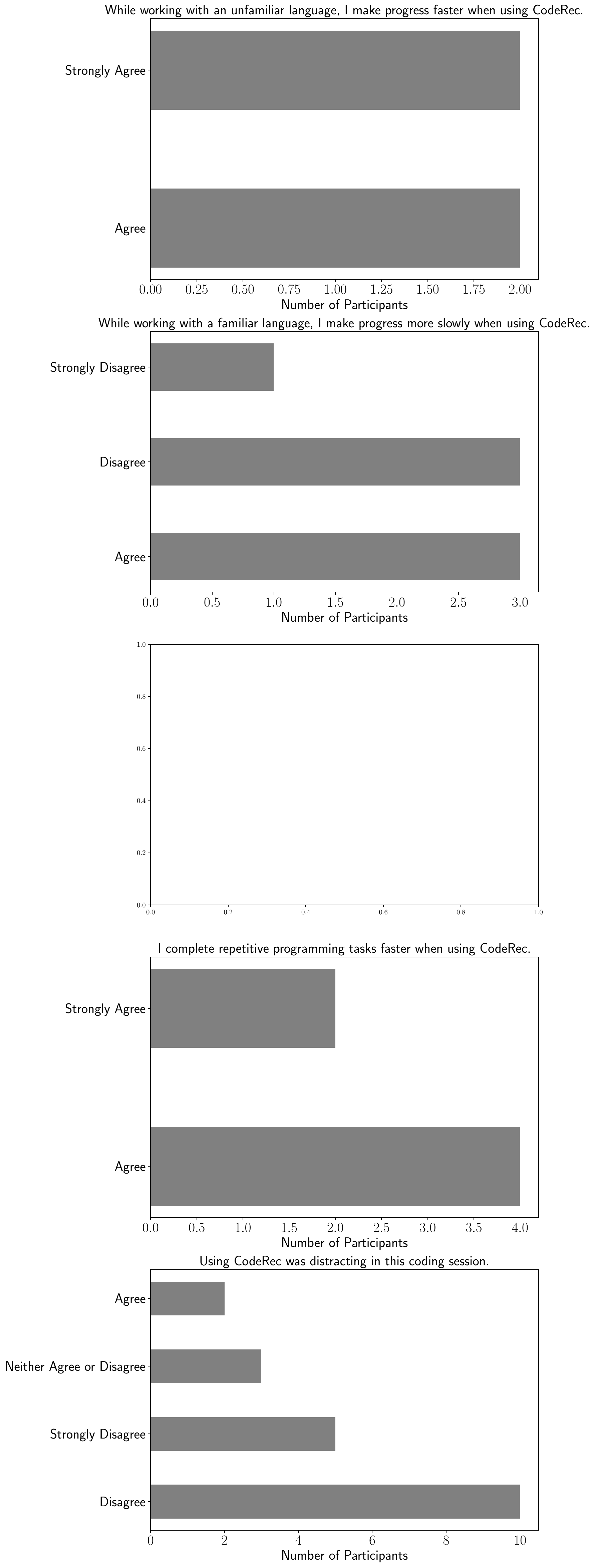}
    \caption{User Study Survey results (4)}
    \label{fig:survey4}
\end{figure}

\subsection{Full User Timelines}

\begin{figure}[H]
    \centering
    \includegraphics[width=\textwidth]{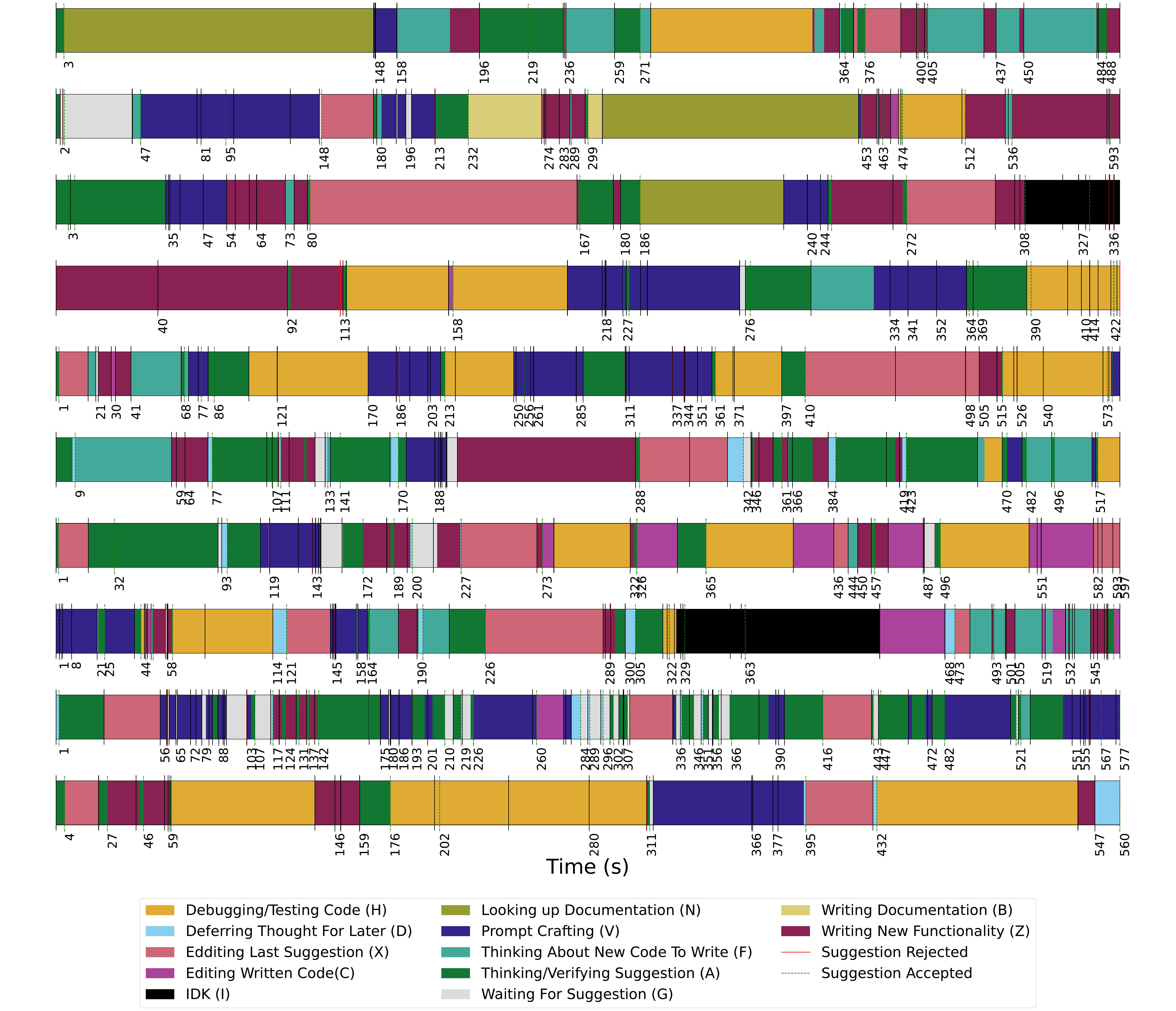}
    \caption{Participants timelines for the first 10 minutes  of their sessions (P1 to P10)  }
    \label{fig:timelines1}
\end{figure}

\begin{figure}[H]
    \centering
    \includegraphics[width=\textwidth]{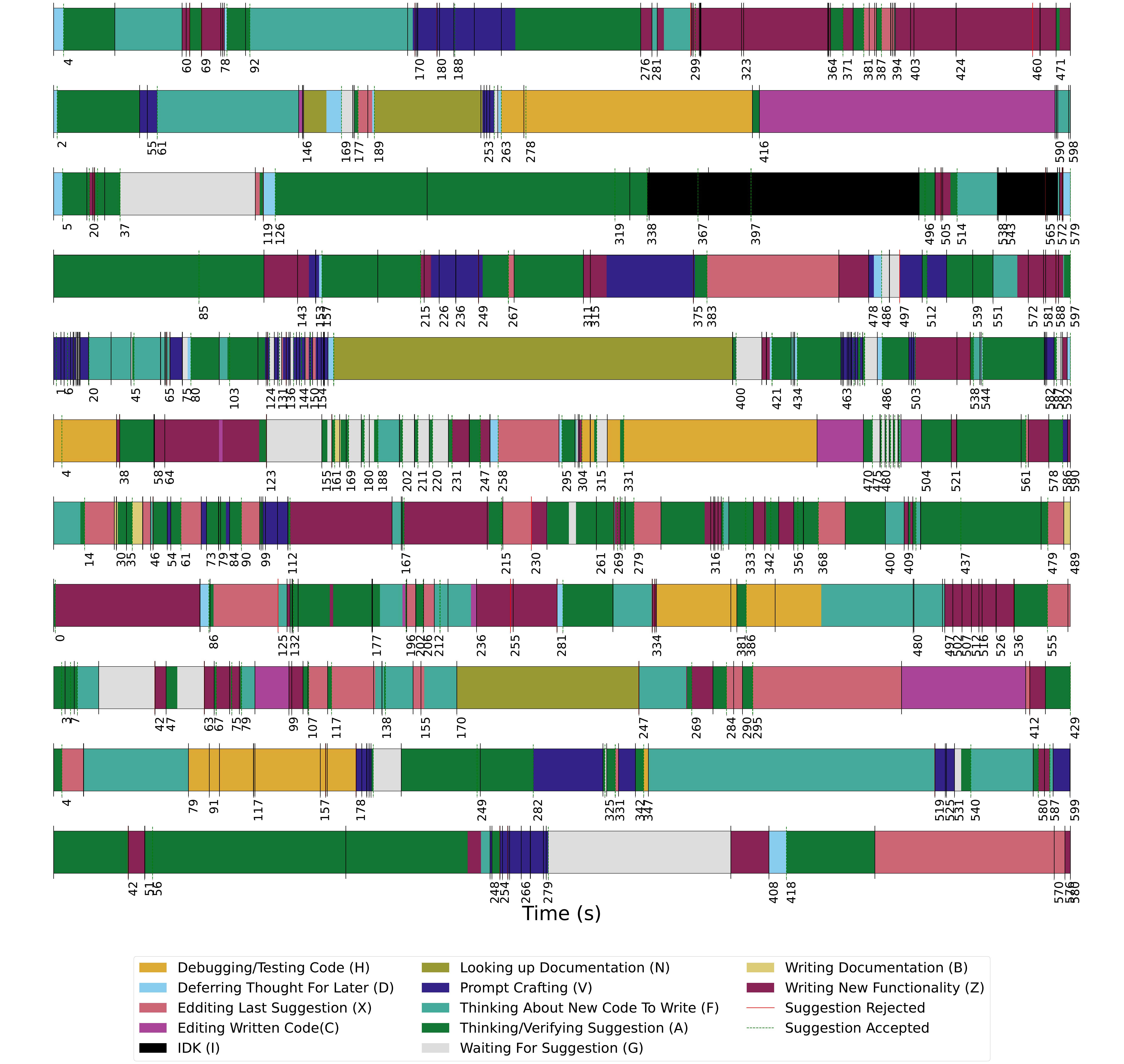}
    \caption{Participants timelines for the first 10 minutes  of their sessions (P11 to P21)  }
    \label{fig:timelines2}
\end{figure}

\subsection{Full \cups Graph}

\begin{figure}[H]
    \centering
    \includegraphics[width=\textwidth]{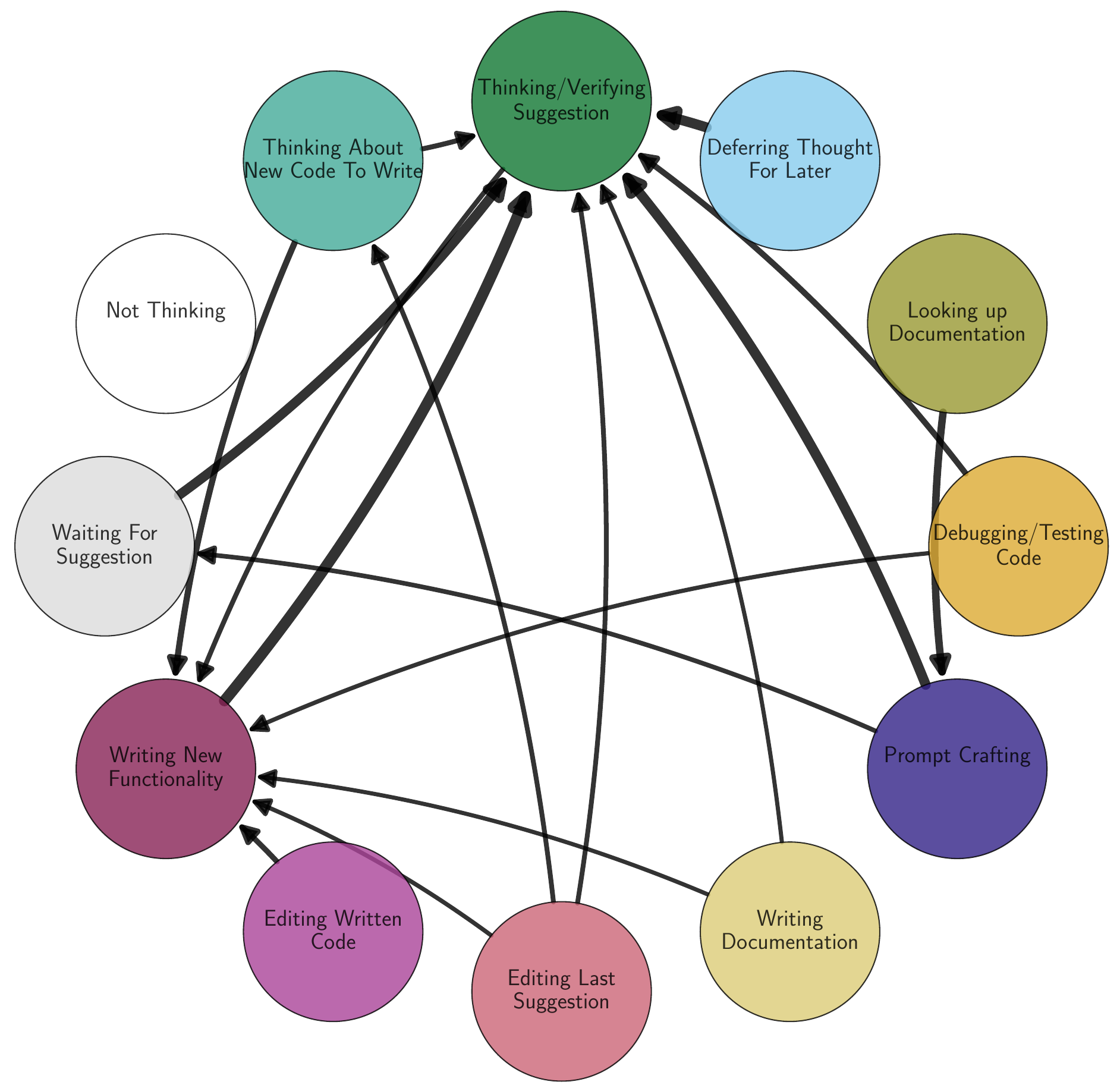}
    \caption{\cups diagram with all transitions shown that occur with probability higher than 0.05  }
    \label{fig:fullgraph}
\end{figure}
\subsection{Probability of Accept by State}

\begin{table}[H]
\caption{Probability of accepting suggestion in next two events  given the user was  in the particular \cups state.}
\begin{tabular}{lc}
    \toprule
\textbf{State} & \textbf{Probability of
Accepting Suggestion} \\ 
\midrule
Thinking/Verifying Suggestion (A)& 0.70\\ 
Prompt Crafting (V)& 0.16\\ 
Looking up Documentation (N)& 0.25\\ 
Writing New Functionality (Z)& 0.19\\ 
Thinking About New Code To Write (F)& 0.21\\ 
Edditing Last Suggestion (X)& 0.16\\ 
Waiting For Suggestion (G)& 0.42\\ 
Editing Written Code(C)& 0.11\\ 
Writing Documentation (B)& 0.36\\ 
Debugging/Testing Code (H)& 0.25\\ 
Deferring Thought For Later (D)& 0.98\\ 
  \bottomrule
\end{tabular}
    \label{tab:prob_Accept_by_state}
\end{table}

\end{document}